\documentclass[]{tPHM2e}

\begin{document}
\doi{10.1080/14786435.20xx.xxxxxx}
\issn{1478-6443}
\issnp{1478-6435}
\jvol{00} \jnum{00} \jyear{2013} 

\markboth{Xiao Lin, Valentin Taufour, Sergey L. Bud'ko and Paul C. Canfield}{Philosophical Magazine}

\articletype{}
\title{Suppression of ferromagnetism in the La(V$_x$Cr$_{1-x}$)Sb$_3$ system}

\author{Xiao Lin$^{\rm a}$, Valentin Taufour$^{\rm a}$, Sergey L. Bud'ko$^{\rm a, b}$, and Paul C. Canfield$^{\rm a, b}$\\\vspace{6pt}  $^{\rm a}${\em{Department of Physics and Astronomy, Iowa State University, Ames, Iowa 50011, U.S.A.}}; $^{\rm b}${\em{Ames Laboratory, U.S. Department of Energy, Iowa State University, Ames, Iowa 50011, U.S.A}}\\\vspace{6pt}}

\maketitle

\begin{abstract}
To explore the possibility of quantum phase transitions and even quantum criticality in LaCrSb$_3$ based compounds, we performed measurements under pressure as well as a vanadium substitution study. The Curie temperature of LaCrSb$_3$ was found to be invariant under pressure. Although pressure was not able to suppress the ferromagnetism, chemical substitution was used as another parameter to tune the magnetism. We grew La(V$_x$Cr$_{1-x}$)Sb$_3$ (\textit{x} = 0 -- 1.0) single crystals, and studied the series by measurements of temperature and field dependent magnetic susceptibility, magnetization, resistivity, and specific heat. Ferromagnetism has been observed for $x \leq 0.22$, and the system manifests a strong anisotropy in its ordered state. The Curie temperature decreases monotonically as the V concentration increases. For $0.42 \leq x \leq 0.73$, the system enters a new magnetic state at low temperatures, and no magnetic ordering above 1.8 K can be observed for $x \geq 0.88$. The effective moment $\mu_{\rm eff}$/Cr varies only slightly as the V concentration increases, from 3.9 $\mu_{\rm B}$ for $x$ = 0 to 2.9 $\mu_{\rm B}$ for $x$ = 0.88. Features related to quantum criticality have not been observed in the La(V$_x$Cr$_{1-x}$)Sb$_3$ system.\bigskip

\bigskip

\end{abstract}

\section{\label{sec:level1a}Introduction}

The study of ferromagnetic materials has long been a focus of research in condensed matter physics. The suppression of an itinerant ferromagnetic transition temperature to zero is of specific interest, since it may lead to the discovery of a quantum critical point (QCP) \cite{Stewart-RMP84, Stewart-RMP01, Stewart-RMP06, Taufour-PRL10, Saxena-Nature00, Huxley-PRB01} which exhibits exotic physical properties, such as non-Fermi liquid behavior and even superconductivity. The Stoner model has been developed to describe a mechanism of an itinerant ferromagnetic system, and is based on the premise that the magnetic properties of the itinerant ferromagnets originate from  de-localized electrons \cite{Stoner-PM33}. In particular these de-localized electrons become part of the conduction band and influence the density of state (DOS) at the Fermi level. Based on the Stoner criterion, $UD(\varepsilon_{\rm F}) >$ 1, where $U$ and $D(\varepsilon_{\rm F})$ are Coulomb repulsion and the DOS at the Fermi level respectively, itinerant ferromagnetism can be suppressed by tuning $U$ and/or $D(\varepsilon_{\rm F})$. The suppression of itinerant ferromagnetism not only results in the decrease of the ordering temperature, but is also accompanied by decrease of the effective and saturated moments per magnetic species. On the other hand, ferromagnetic ordering can also arise from the interactions of local magnetic moments \cite{Lacheisserie05, Ashcroft-76}. As the exchange interaction favors parallel spin alignments, the materials show spontaneous magnetization. Suppressing the ferromagnetism by diluting the local moments, does not reduce the size of effective or saturated moments (per moment bearing ion) and, does not necessarily lead to a QCP \cite{Jia-NP07, Szytula94}. New magnetic states, such as spin glass, may also emerge in the diluted magnetic system \cite{Mydosh-93, Wiener-PRB00}. Therefore, the suppression of ferromagnetism by substitution may offer an opportunity to approach a QCP, or may result a glassy state, and in doing so, sheds light onto the nature of the ordering mechanism of a specific ferromagnetic system.

LaCrSb$_3$ has been reported to order ferromagnetically below $T_{\rm C} \sim$ 125 -- 142 K, with the differences arising from the sample preparation methods \cite{Hartjes-JMMM97, Leonard-JAP99, Leonard-JAC00, Raju-CM98, Jackson-PRB01}. LaCrSb$_3$ crystallizes in an orthorhombic structure (space group $Pbcm$), where Cr occupies one single crystallographic site 4$c$ \cite{Brylak-ZNB95}. Extensive investigations into LaCrSb$_3$ have been undertaken, and the compound is found to have a rich magnetic phase diagram \cite{Jackson-PRB01, Granado-PRL02, MacFarlane-PB06, Crerar-JSSC12, Choi-JAP07}. LaCrSb$_3$ exhibits unconventional magnetic behavior with a canted ferromagnetism in $bc$-plane. A spin-reorientation transition can be observed in the $bc$-plane at $\sim$ 95 K, and can be suppressed  by a small applied magnetic field $\sim$ 250 Oe \cite{Jackson-PRB01}. Whereas some studies claim LaCrSb$_3$ is an itinerant ferromagnet \cite{Raju-CM98, Jackson-PRB01}, the nature of its magnetic moments is still under debate. A neutron scattering study suggests a coexistence of localized and itinerant spins in LaCrSb$_3$ \cite{Granado-PRL02}. As La is not moment-bearing, the Cr ion plays the primary role in the magnetism of LaCrSb$_3$. Band structure calculations and the X-ray photoelectron spectroscopy studies find that the 3$d$ electrons of the Cr exhibit a large DOS peak at/near the Fermi level in the paramagnetic state \cite{Crerar-JSSC12, Choi-JAP07, Richter-JMMM04}, which, based on the Stoner criterion, has a possibility of inducing the ferromagnetic instability. 

An itinerant magnetic system can often be perturbed by applying pressure or via chemical substitution. Take MnSi \cite{Thessieu-SSC95}, UGe$_2$ \cite{Taufour-PRL10} and La(V$_x$Cr$_{1-x}$)Ge$_3$ \cite{Lin-Ge} as examples; in each, the ferromagnetic state disappears as pressure is applied. Thus, pressure might be able to suppress the ferromagnetic phase and lead to a QCP or quantum phase transition (QPT) in LaCrSb$_3$. Also the DOS can be changed by chemical substitutions for the Cr atoms. LaVSb$_3$, which is an isostructural compound to LaCrSb$_3$, has no magnetic ordering down to 2 K \cite{Brylak-ZNB95, Jackson-PRB01, Sefat-JMMM08}. It is found that the Fermi level in LaVSb$_3$ is shifted away from the highest peak of the DOS \cite{Choi-JAP07}. Thus, the ferromagnetism in LaCrSb$_3$ may also be suppressed by substituting V for Cr atom.

Previous work on polycrystalline samples showed that V substitution does suppress the ferromagnetic transitions, and claimed that the mechanism of the ferromagnetic ordering can not be explained by a simple localized magnetic moment model \cite{Dubenko-JAP01}. Only the temperature dependence of magnetization was measured on the V-doped polycrystalline samples. The nominal V substitution reached only up to 20$\%$, and the precise stoichiometry of this doped system was not analyzed experimentally. It is also not clear at which concentration the ferromagnetism was fully suppressed. In order to better understand the effects of V substitution on the magnetic state of this system, detailed measurements of the transport and thermodynamic properties of systematically substituted single crystals are necessary.

In this work, we report the synthesis of single crystalline La(V$_x$Cr$_{1-x}$)Sb$_3$ (\textit{x} = 0 -- 1.0) samples, and present a systematic study of their transport and thermodynamic properties. In addition, measurements of magnetization under pressure were performed on the LaCrSb$_3$ sample. Whereas the Curie temperature is essentially invariant under pressure, the ferromagnetic ordering is systematically suppressed as the V concentration increases from $x$ = 0 to $x$ = 0.36. For $0.42 \leq x \leq 0.73$, the system enters into a new magnetic ground state, possibly a complex glassy state. For even higher V-doped compounds, $ x \geq 0.88$, the samples stay in the paramagnetic state down to 2 K. The magnetic anisotropy also changes with the V substitution. Although the effective moment per Cr varies slightly as the V concentration increases, possibly suggesting a valence change of Cr ion induced by V substitution, there is no indication of $\mu_{\rm eff}$ decreasing toward zero and the Cr moment appears to be robust and fundamentally local-moment like in nature. No experimental features expected in the vicinity of a QCP have been observed by either applied pressure or chemical substitution. 

\section{\label{sec:level1b}Experimental Details}

\begin{figure}
\begin{center}
\resizebox*{8.5cm}{!}{\includegraphics{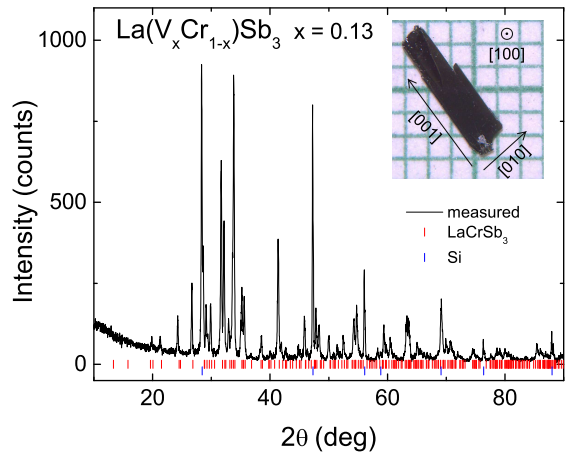}}
\caption{\label{fig:Sb-X-ray} Powder X-ray diffraction pattern of La(V$_{x}$Cr$_{1-x}$)Sb$_3$ ($x$ = 0.13). Inset: photo of a single crystalline sample ($x$ = 0.06) on a millimeter grid.}
\end{center}
\end{figure}

Single crystalline La(V$_x$Cr$_{1-x}$)Sb$_3$ samples were synthesized via high-temperature solution method with excess Sb as self-flux \cite{Jackson-PRB01, Sefat-JMMM08, Canfield-PMB92, Canfield-Euroschool10}. High purity ($>$ 3N) elements with the starting stoichiometry of La : V : Cr : Sb = 8 : $x$ : 8-$x$ : 84, were placed in a 2 mL alumina crucible and sealed in a fused silica tube under a partial pressure of high purity argon gas. The ampoule containing the growth materials was heated up to 1180 $^\circ$C over 3 h and held at 1180 $^\circ$C for another 3 h. The growth was then cooled to 750 $^\circ$C over $\sim$ 85 h at which temperature the excess liquid was decanted using a centrifuge \cite{Canfield-PMB92, Canfield-Euroschool10}. Single crystals of La(V$_x$Cr$_{1-x}$)Sb$_3$ grew as rectangular plates, with shiny surfaces that had a few drops of residual Sb-rich flux on them. An example of such a crystal is shown in the inset of fig \ref{fig:Sb-X-ray}. The sizes of crystals increase as the V-concentration increases, varying from $\sim$ 3.5 $\times$ 1.5 $\times$ 0.7 mm$^3$ for LaCrSb$_3$ to being crucible limited, $\sim$ 8 $\times$ 6 $\times$ 2 mm$^3$ for LaVSb$_3$.

Powder X-ray diffraction data were collected at room temperature on a Rigaku MiniFlex II diffractometer with Cu K$\alpha$ radiation. Samples with rod-like shape were selected for measurement. Data collection was performed with the counting time of 2 s for every 0.02 degree. The refinement was conducted using the program Rietica \cite{Rietica}. Error bars associated with the values of the lattice parameters were determined by statistical errors, and a Si powder standard was used as an internal reference. To identify the crystallographic orientation, real-time back-scattering Laue diffraction measurements were performed with Mo source ($\lambda \sim 0.7093 \r{A}$). The structural solutions were refined by the Cologne Laue Indexation Program \cite{Laue}.

\begin{figure}
\begin{center}
\resizebox*{8.5cm}{!}{\includegraphics{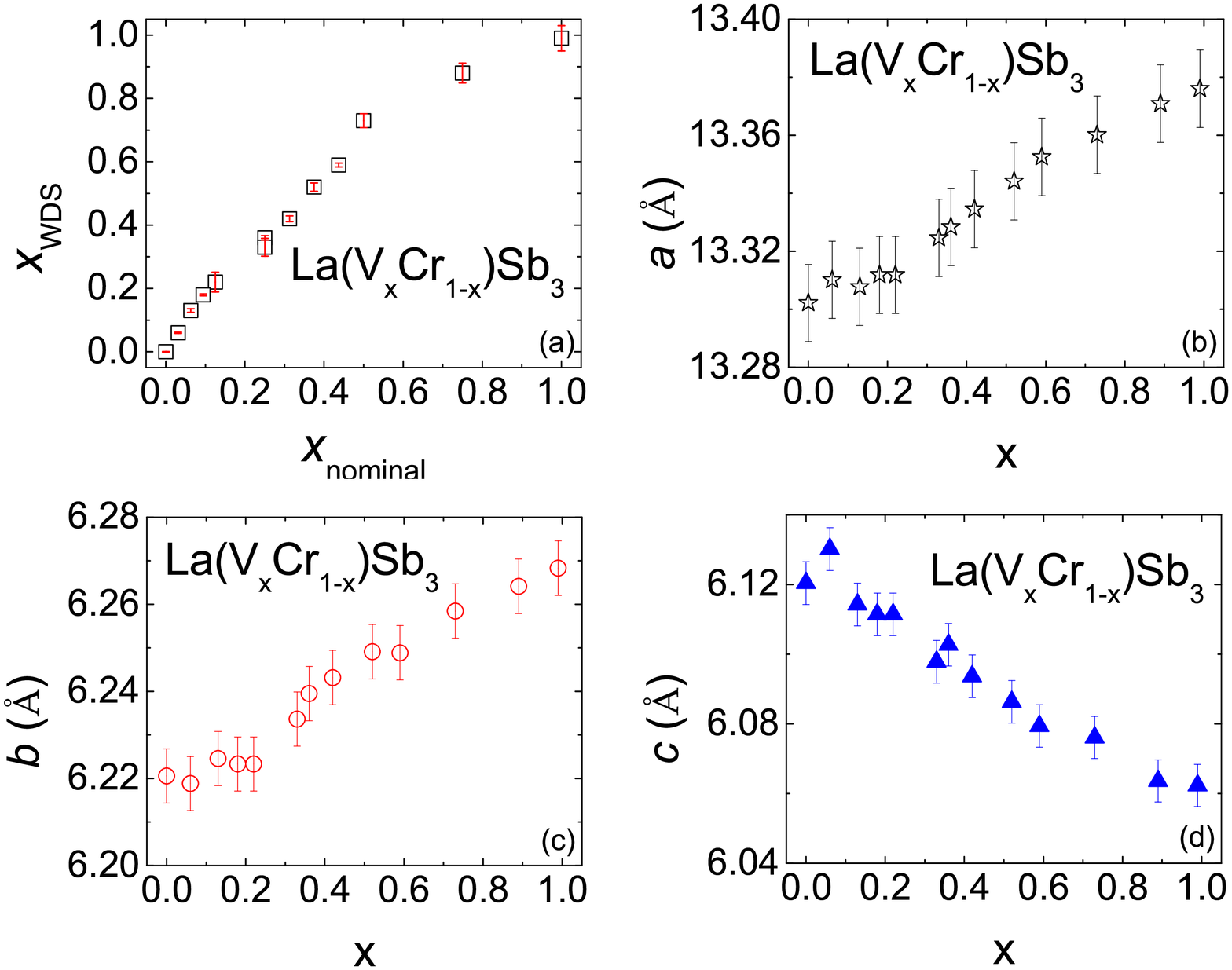}}
\caption{\label{fig:Sb-Lattice} (a) $x_{\rm WDS}$ vs. $x_{\rm nominal}$. The lattice parameters of single crystalline La(V$_x$Cr$_{1-x}$)Sb$_3$ compounds as a function of the V-concentration are shown in (b) $a$ vs. $x$, (c) $b$ vs. $x$, and (d) $c$ vs. $x$.  Note: the measured, $x_{\rm WDS}$ were used to indicate the composition of the compounds in this series.}
\end{center}
\end{figure}

Elemental analysis of the samples was performed using wavelength-dispersive X-ray spectroscopy (WDS) in a JEOL JXA-8200 electron probe microanalyzer. Only clear and shiny surface regions were selected for determination of the sample stoichiometry, i.e. regions with residual Sb flux droplets were avoided. For each composition, the WDS data were collected from multiple points on the same sample. 

Measurements of field and temperature dependent magnetization were performed in a Quantum Design, Magnetic Property Measurement System (MPMS) superconducting quantum interference device (SQUID) magnetometer. The ac resistivity was measured by a standard four-probe method in a Quantum Design, Physical Property Measurement System (PPMS). Samples were polished into long rectangular bars. Platinum wires were attached to the sample using Epo-tek H20E silver epoxy, with the current flowing along the $c$-axis. The absolute values of resistivity are accurate to $\pm 15\%$ due to the accuracy of measurements of electrical contacts' positions. The residual resistivity ratio is defined as RRR = $\rho$ (300 K)/$\rho$ (2.0 K). 

Temperature dependent specific heat data were measured in the PPMS using the relaxation technique in zero field for representative samples. The specific heat of LaVSb$_3$ was used to estimate the non-magnetic contribution to the specific heat of LaCrSb$_3$. The magnetic contribution to specific heat from the Cr ions was calculated by the relation of $C_{\rm M}$ = $C_{\rm p}$(LaCrSb$_3$)-$C_{\rm p}$(LaVSb$_3$).  

The temperature dependent field-cooled magnetization of a single crystal under pressure was measured in the MPMS magnetometer in a magnetic field of 100 Oe applied along the c-axis. Pressures of up to 5.3 GPa were achieved with a moissanite anvil cell \cite{Alireza-07}. The body of the cell is made of Cu-Ti alloy and the gasket is made of Cu-Be. Daphne 7474 was used as a pressure transmitting medium \cite{Murata-RSI08}, and the pressure was determined at 77 K by the ruby fluorescence technique.

\section{\label{sec:level1c}Results and Analysis}
\subsection{\label{sec:level2a}Crystal Stoichiometry and Structure}

\begin{table}
  \tbl{The WDS data for La(V$_x$Cr$_{1-x}$)Sb$_3$. $N$ is the number of points measured on one sample, $x_{\rm nominal}$ is the nominal concentration, $x_{\rm WDS}$ is the average $x$ value measured, and 2$\sigma$ is two times the standard deviation of the $N$ values measured.}
{\begin{tabular}{@{}cccccccccccccc}\toprule
$N$	&12&	12	&	12	&	12	&	12	&	12	&	12	&	12	&	12	&	12	&	12	&	12&12\\
\colrule
  $x_{\rm nominal}$	&0& 0.03 &	0.06	&	0.09	&	0.13	&	0.25	&	0.25	&	0.31	&	0.38	&	0.44	&	0.50	&	0.75&1.0\\
$x_{\rm WDS}$	&0&0.06	&	0.13	&	0.18	&	0.22	&	0.33	&	0.36	&	0.42	&	0.52	&	0.59	&	0.73	&	0.88&1.0\\
2$\sigma$	&0.01& 	0.01	&	0.01	&	0.01	&	0.06	&	0.06	&	0.01	&	0.02	&	0.03	&	0.01	&	0.04	&	0.06&0.01\\
   \botrule
  \end{tabular}}
\label{tab:Sb-WDS}
\end{table}

The stoichiometry of the La(V$_x$Cr$_{1-x}$)Sb$_3$ samples was inferred from WDS measurements. Table \ref{tab:Sb-WDS} summarizes the atomic percent of each element determined from the weight percent obtained from the analyses. The error bar is taken as twice the standard deviation $\sigma$. As shown in fig. \ref{fig:Sb-Lattice} (a), the actual V-concentration $x_{\rm WDS}$ follows the initial stoichiometry $x_{\rm nominal}$ systematically, ranging from 0 to 1, and the small 2$\sigma$-value suggests that the samples are homogeneous, at least on the length scale probed by the WDS measurements $\sim$ 1 $\mu$m. In the following, the measured, $x_{\rm WDS}$, rather than $x_{\rm nominal}$ values will be used to indicate the composition of the compounds in this series.

The crystal structure and orientation were confirmed by back-scattering Laue diffraction. Consistent with the reported data \cite{Brylak-ZNB95, Jackson-PRB01, Sefat-JMMM08}, this series of compounds form in an orthorhombic structure, $Pbcm$ (No. 57). As shown in the inset of fig. \ref{fig:Sb-X-ray}, the $a$-axis was verified to be perpendicular to the rectangular plate, and the $c$-axis is parallel to the longest side, consistent with the reported data \cite{Sefat-JMMM08}. Powder X-ray diffraction patterns were collected on ground single crystals from each compound. Fig. \ref{fig:Sb-X-ray} gives powder X-ray diffraction pattern for $x$ = 0.13 as an example. The main phase can be refined with LaCrSb$_3$'s reflection pattern ($Pbcm$ structure), consistent with the Laue diffraction. No clear trace of Sb residue or other secondary solidification can be detected, and similar results ($Pbcm$ structure) were obtained for the rest of the series. The lattice parameters obtained by the analysis of the powder X-ray diffraction data are presented in fig. \ref{fig:Sb-Lattice} (b) -- (d). The lattice parameters $a$, $b$ and $c$ all manifest systematic changes as the $x$ increases, which is consistent with the reported data \cite{Brylak-ZNB95, Jackson-PRB01}. Crystallographically, the transition metal elements in LaCrSb$_3$ and LaVSb$_3$ occupy the same unique site $4c$ \cite{Brylak-ZNB95}. 

\subsection{\label{sec:level2b}Physical properties of La(V$_x$Cr$_{1-x}$)Sb$_3$ ($x$ = 0 and 1.0)}

\begin{figure}
\begin{center}
\resizebox*{8.5cm}{!}{\includegraphics{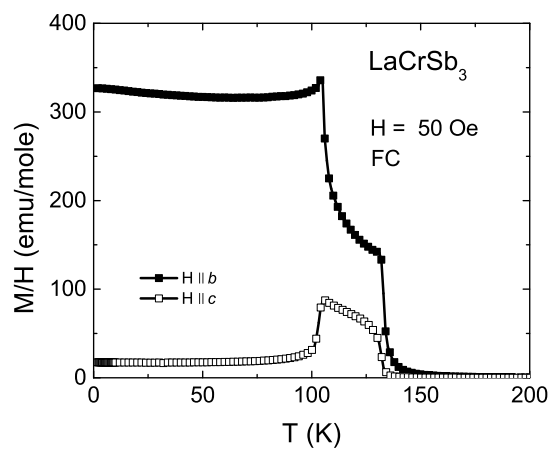}}
\caption{\label{fig:Sb-Cr_MT} Anisotropic field-cooled (FC) magnetization as a function of temperature for LaCrSb$_3$ at 50 Oe.}
\end{center}
\end{figure}

The anisotropic, temperature-dependent, field-cooled (FC) magnetization of LaCrSb$_3$ is shown in fig. \ref{fig:Sb-Cr_MT}. The measurements were performed with the applied field parallel to $b$- and $c$-axes at 50 Oe. As is shown, the magnetization rises sharply near 130 K for both \textbf{H} $\parallel b$ and \textbf{H} $\parallel c$, indicating a transition to a low-temperature ferromagnetic state. A second anomaly can be observed in both directions at around 100 K, which can be associated with spin reorientation, as suggested by previous studies \cite{Jackson-PRB01, Granado-PRL02}. Below roughly look, the magnetization data in both directions remain almost constant as temperature is lowered.

\begin{figure}
\begin{center}
\resizebox*{8.5cm}{!}{\includegraphics{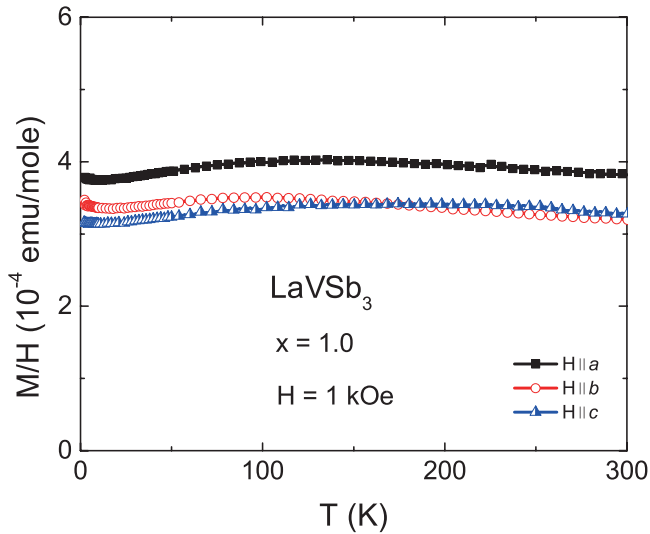}}
\caption{\label{fig:Sb-V_MT} Anisotropic magnetic susceptibility as a function of temperature of LaVSb$_3$ measured at 1 kOe.}
\end{center}
\end{figure}

The anisotropic magnetic susceptibility of LaVSb$_3$ was measured at 1 kOe, as shown in fig. \ref{fig:Sb-V_MT}. It is clear that the magnetic susceptibility has weakly positive values in all three directions, and $M(T)/H$ is essentially temperature independent. It is evident that LaVSb$_3$ follows Pauli magnetic behavior, and is consistent with the reported data \cite{Jackson-PRB01, Sefat-JMMM08}.

\begin{figure}
\begin{center}
\resizebox*{8.5cm}{!}{\includegraphics{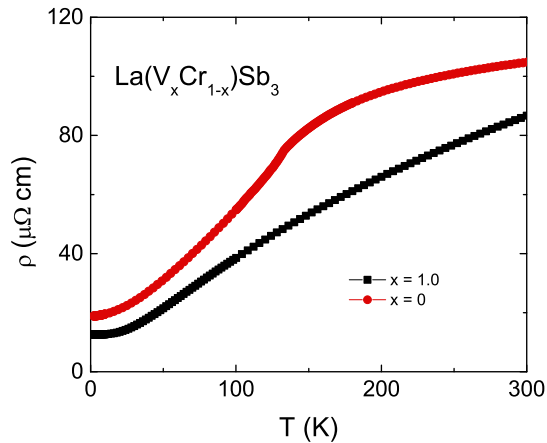}}
\caption{\label{fig:Sb-Cr_RT} The electrical resistivity $\rho$ as a function of temperature for La(V$_x$Cr$_{1-x}$)Sb$_3$ ($x$ = 0 and 1.0).}
\end{center}
\end{figure}

Figure \ref{fig:Sb-Cr_RT} presents the electrical resistivity data of La(V$_x$Cr$_{1-x}$)Sb$_3$ ($x$ = 0 and 1.0) as a function of temperature. To within 15$\%$, the room temperature resistivity values $\rho$ (300 K) are about 105 $\mu\Omega$ cm for $x = 0$ and 87 $\mu\Omega$ cm for $x = 1.0$. At high temperatures, the electrical resistivity decreases linearly upon cooling, characteristic of normal metallic behavior. For $x = 0$, a dramatic anomaly occurs at about 132 K, which is most likely due to the loss of spin disorder scattering and can be associated with the ferromagnetic transition. For $x = 1.0$, no anomaly was observed for temperatures above 1.8 K.

\begin{figure}
\begin{center}
\resizebox*{8.5cm}{!}{\includegraphics{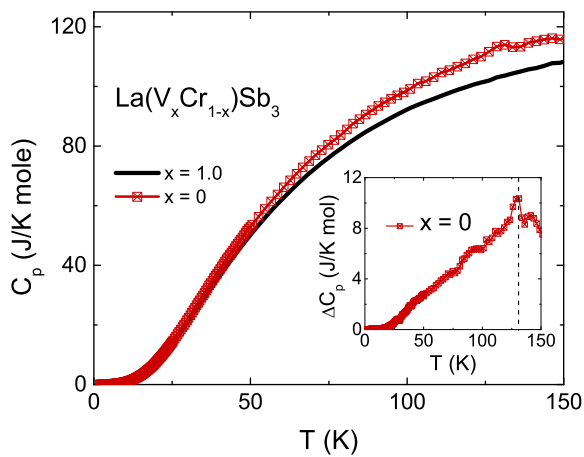}}
\caption{\label{fig:Sb-Cr_CT} Temperature-dependent of specific heat of La(V$_x$Cr$_{1-x}$)Sb$_3$ ($x$ = 0 and 1). Inset: magnetic contributions to the specific heat $\bigtriangleup C_{\rm p}$ as a function of temperature for $x$ = 0. The dashed line indicates the ordering temperature.}
\end{center}
\end{figure}

The temperature-dependent specific heat data for the La(V$_x$Cr$_{1-x}$)Sb$_3$ ($x$ = 0 and 1) are presented in fig. \ref{fig:Sb-Cr_CT}. The specific heat can be estimated by the relation $C_{\rm p}(T)$ = $C_{\rm e}$ + $C_{\rm ph}$ + $C_{\rm M}$, where $C_{\rm e}$ is the conduction electron contribution, $C_{\rm ph}$ is the phonon contribution, and $C_{\rm M}$ is the magnetic contribution. $C_{\rm e}$ + $C_{\rm ph}$ can be roughly approximated by the $C_{\rm p}$ data of LaVSb$_{3}$. Thus, the magnetic contribution $C_{\rm M}$ was, to the specific neat of LaCrSb$_3$, evaluated as $\bigtriangleup C_{\rm p}$ = $C_{\rm p}$(LaCrSb$_3$)-$C_{\rm p}$(LaVSb$_3$). A cusp can be seen in $\bigtriangleup C_{\rm p}$, as shown in the inset of fig.\ref{fig:Sb-Cr_CT}. This is the first time that the ferromagnetic transition of LaCrSb$_3$ has been observed in the specific heat data. This anomaly can be associated with the ferromagnetic transition. The ordering temperature $T_{\rm C}$ obtained from $\bigtriangleup C_{\rm p}$ data for $x$ = 0 is about 132 K, as indicated by the dash line in the inset of fig. \ref{fig:Sb-Cr_CT}. 

\subsection{\label{sec:level2c}Effects of pressure on the magnetic properties of LaCrSb$_3$}

\begin{figure}
\begin{center}
\resizebox*{8.5cm}{!}{\includegraphics{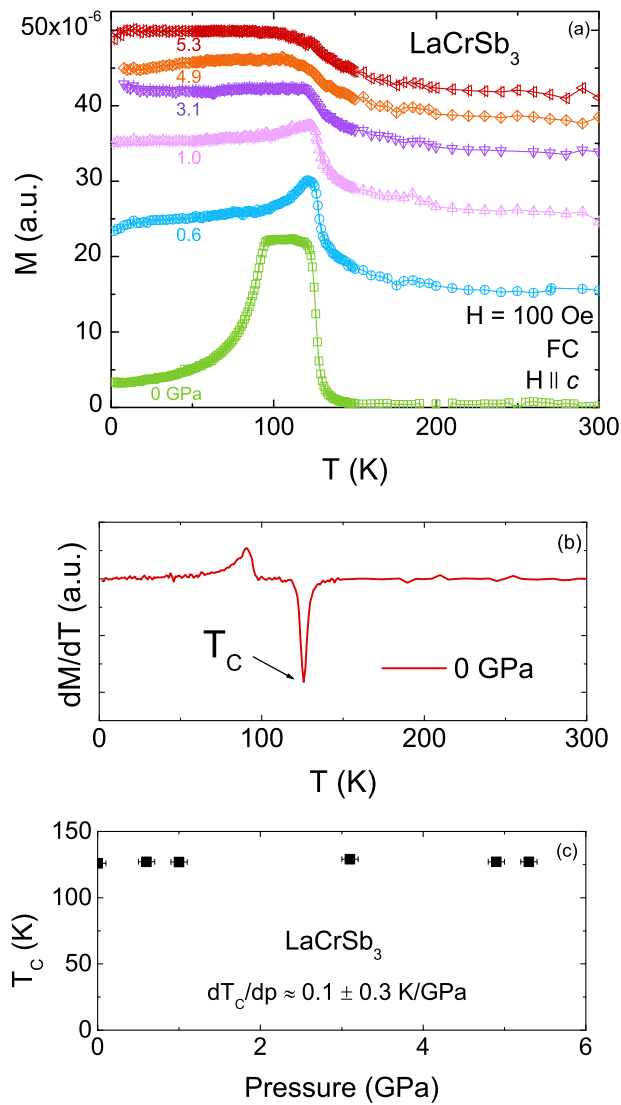}}
\caption{\label{fig:Sb-Pressure} (a) Temperature dependence of the FC magnetization for LaCrSb$_3$ under different pressures with \textit{H} = 100 Oe, \textbf{H} $\parallel c$. Note: The values of the magnetization are shifted for clarity. (b) d$M$/d$T$ vs. $T$: the arrow indicates the Curie temperature. (c) Pressure dependence of $T_{\rm C}$ for LaCrSb$_3$, where $T_{\rm C}$ is determined by the minimum point in $dM(T)/dT$.}
\end{center}
\end{figure}

In an attempt to suppress the ferromagnetism in LaCrSb$_3$, hydrostatic pressures up to 5.3 GPa were applied. Figure \ref{fig:Sb-Pressure} (a) shows the temperature dependence of the filed-cooled magnetization of LaCrSb$_3$ under different pressures. At lower pressures, the ferromagnetic transition is revealed by a rather sharp increase of the magnetization. Defined here as a minimum point in $dM(T)/dT$ (as seen in fig. \ref{fig:Sb-Pressure} (b)), the Curie temperature, $T_{\rm C}$, is plotted as a function of applied pressure in fig. \ref{fig:Sb-Pressure} (c). $T_{\rm C}$ changes only very slightly with applied pressure with $dT_{\rm C}/dp \approx 0.1 \pm 0.3$ K/GPa, suggesting that the ferromagnetism is robust with respect to pressure, at least up to 5.3 GPa. At ambient pressure, the spin reorientation is seen as a sharp decrease of the field-cooled magnetization measured along the $c$-axis (fig. \ref{fig:Sb-Pressure} (a) and fig. \ref{fig:Sb-Cr_MT}). A decrease of the magnetization is still observed for pressure of 0.6 and 1 GPa, although the plateau can not be observed. The decrease of magnetization cannot be detected above 3 GPa.

\subsection{\label{sec:level2d}Effects of chemical substitution on the physical properties of La(V$_x$Cr$_{1-x}$)Sb$_3$}


\begin{figure}
\begin{center}
\resizebox*{8.5cm}{!}{\includegraphics{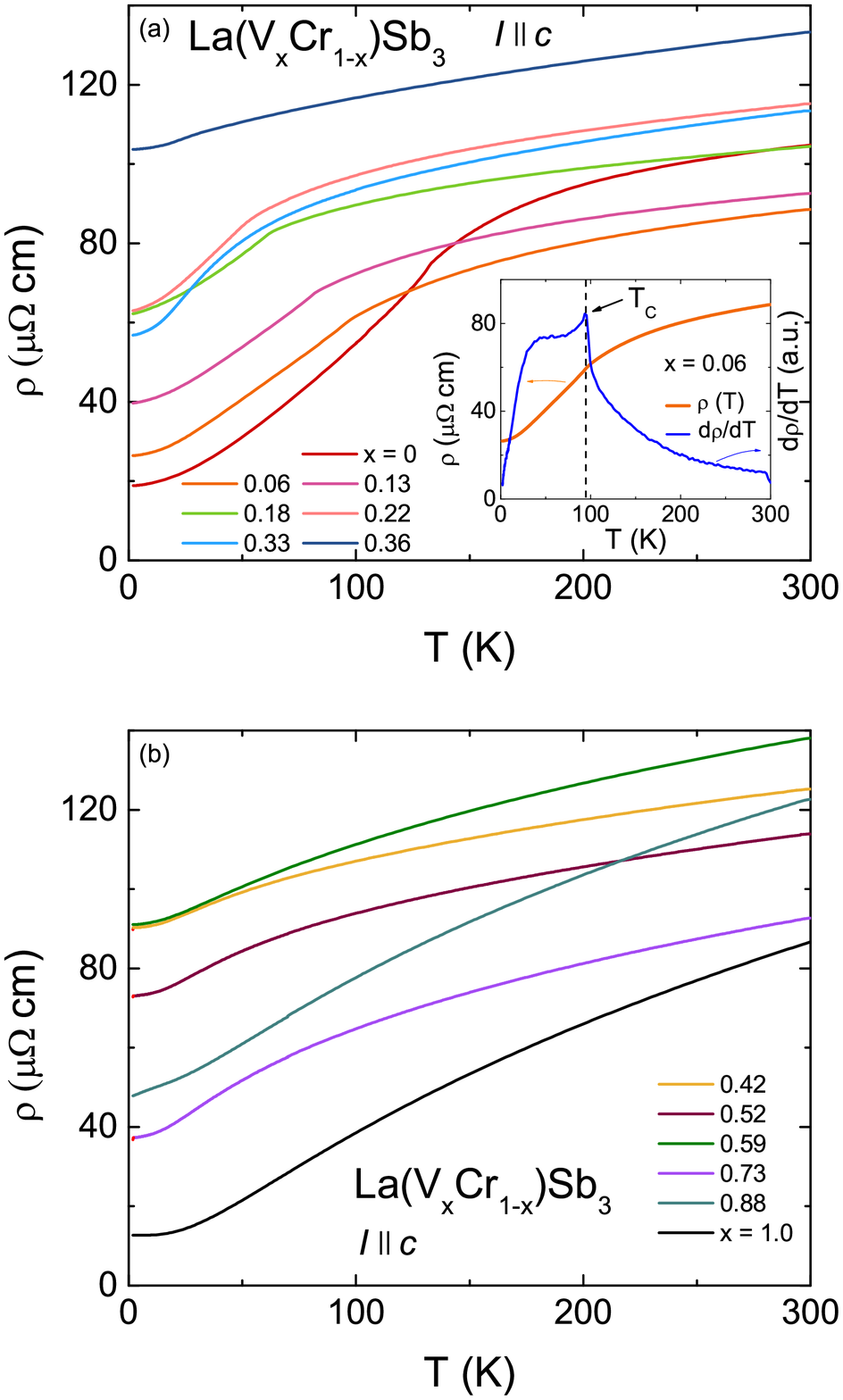}}
\caption{\label{fig:Sb-RT} The electrical resistivity $\rho$ as a function of temperature for La(V$_x$Cr$_{1-x}$)Sb$_3$. (a) $x$ = 0 -- 0.36. Inset: $\rho (T)$ and $d\rho/dT$ for $x$ = 0.06. The arrow indicates the criterion used to determine the Curie temperature $T_{\rm C}$. (b) $x$ = 0.42 -- 1.0.}
\end{center}
\end{figure}

Given that accessible pressures appear to have little or no effect on the ferromagnetic transition temperature of LaCrSb$_3$, we decided to study the effects of chemical substitution on the physical properties of the La(V$_x$Cr$_{1-x}$)Sb$_3$ series. The electrical resistivity data, as a function of temperature for La(V$_x$Cr$_{1-x}$)Sb$_3$ are presented in fig. \ref{fig:Sb-RT} (a) and (b). The room temperature resistivity values $\rho$ (300 K) of all compounds are in the range of 80 -- 140 $\mu\Omega$ cm. For low $x$-value samples, the distinct drop in the resistivity below 150 K is probably associated with the ferromagnetic transition. This anomaly moves to lower temperatures and is broadened as the V concentration increases up to 0.36 (seen in fig. \ref{fig:Sb-RT} (a)). For $x \geq$ 0.42, this feature can no longer be clearly observed, as shown in fig. \ref{fig:Sb-RT} (b). The inset of fig. \ref{fig:Sb-RT} (a) provides the criterion used to infer Curie temperature $T_{\rm C}$ --- the peak position in the $d\rho/dT$ indicated by the arrow, and the inferred $T_{\rm C}$ values are summarized in Table \ref{tab:Sb-Temperature} (below). With current along $c$-axis, samples for $x$ = 1.0 and 0 have RRR of $\simeq$ 6.8 and 5.6, respectively. The lower RRR values for the intermediate $x$-value compounds are due to increased site disorder caused by the substitution.

\begin{figure}
\begin{center}
\resizebox*{8.5cm}{!}{\includegraphics{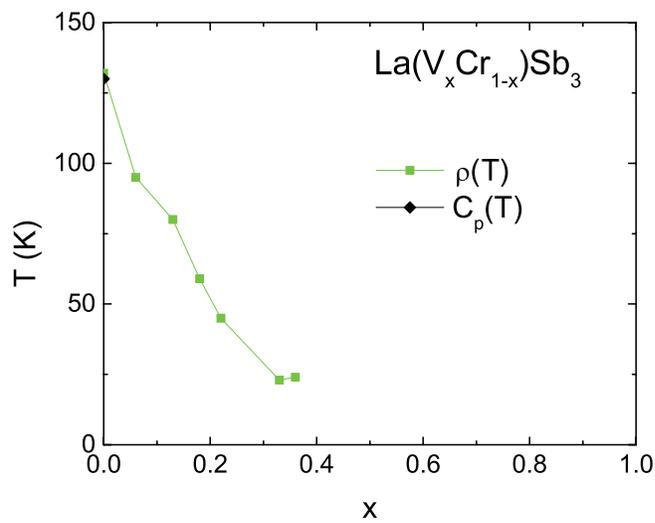}}
\caption{\label{fig:Sb-R_PD} $x$-dependent transition temperatures for La(V$_x$Cr$_{1-x}$)Sb$_3$ determined by $\rho(T)$ and $C_{\rm p}(T)$ measurements.}
\end{center}
\end{figure}

Based on the resistivity and specific heat data, a $T$ -- $x$ phase diagram was assembled. As shown in fig. \ref{fig:Sb-R_PD}, the Curie temperature decreases systematically as the V concentration increases. For lower V-doped compounds, $x < $ 0.36, the system possess a paramagnetic state at high temperatures, and transits into a magnetically ordered state at low temperatures. For higher V-doped compounds, $x \geqslant$ 0.42, as no feature can be observed in the resistivity data, the magnetic state in this region is not clear.

\begin{figure}
\begin{center}
\resizebox*{8.5cm}{!}{\includegraphics{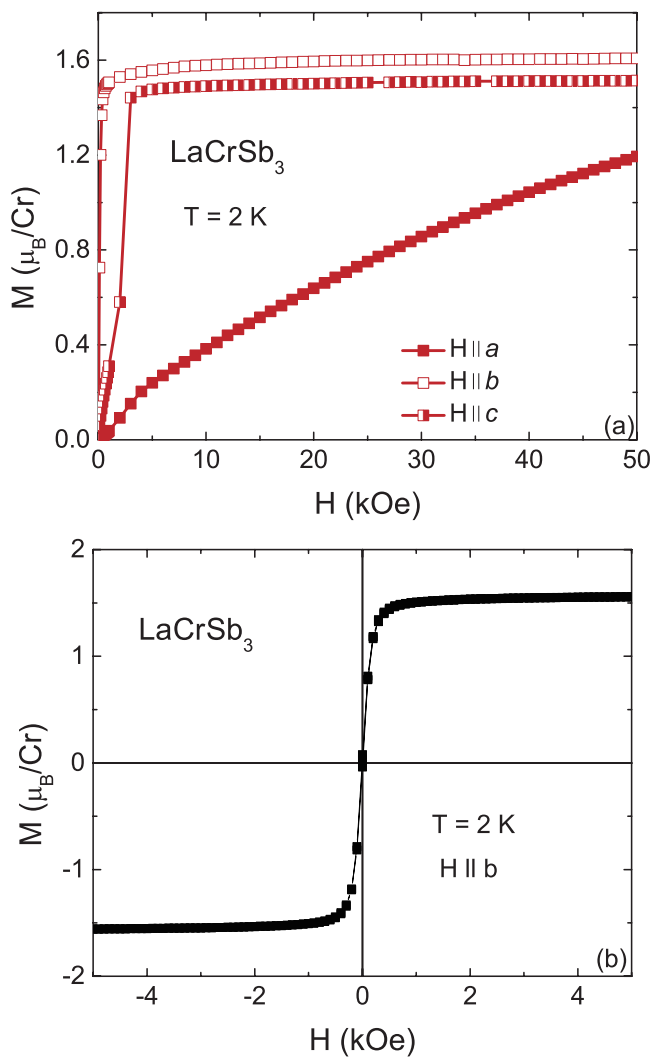}}
\caption{\label{fig:Sb-MH_0} (a) Zero-field-cooled (ZFC) anisotropic field-dependent magnetization isotherms taken at 2 K. (b) Hysteresis loop measured at 2 K with \textbf{H} $\parallel b$ for the LaCrSb$_3$. The applied magnetic field changes from zero to 50 kOe, to -50 kOe, and then up to 50 kOe again.}
\end{center}
\end{figure}

\begin{figure}
\begin{center}
\resizebox*{8.5cm}{!}{\includegraphics{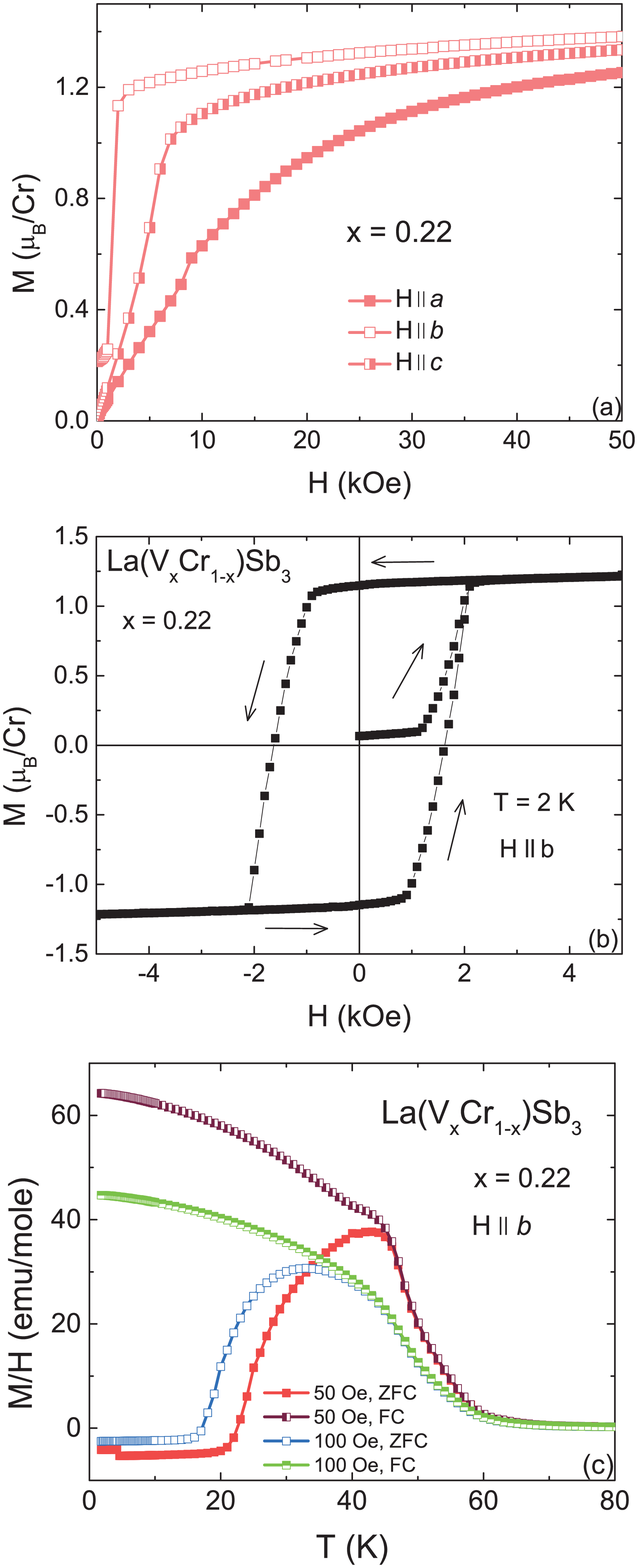}}
\caption{\label{fig:Sb-MH_22} (a) Zero-field-cooled (ZFC) anisotropic field-dependent magnetization isotherms taken at 2 K. (b) Hysteresis loop measured at 2 K with \textbf{H} $\parallel b$. The applied magnetic field changes from zero to 50 kOe, to -50 kOe, and then up to 50 kOe again. The arrows indicate the directions of the field sweeping. (c) Zero-field-cooled (ZFC) and field-cooled (FC) temperature dependence of the magnetic susceptibility taken at 50 and 100 Oe with \textbf{H} $\parallel b$ for the La(V$_x$Cr$_{1-x}$)Sb$_3$ ($x$ = 0.22).}
\end{center}
\end{figure}

\begin{figure}
\begin{center}
\resizebox*{8.5cm}{!}{\includegraphics{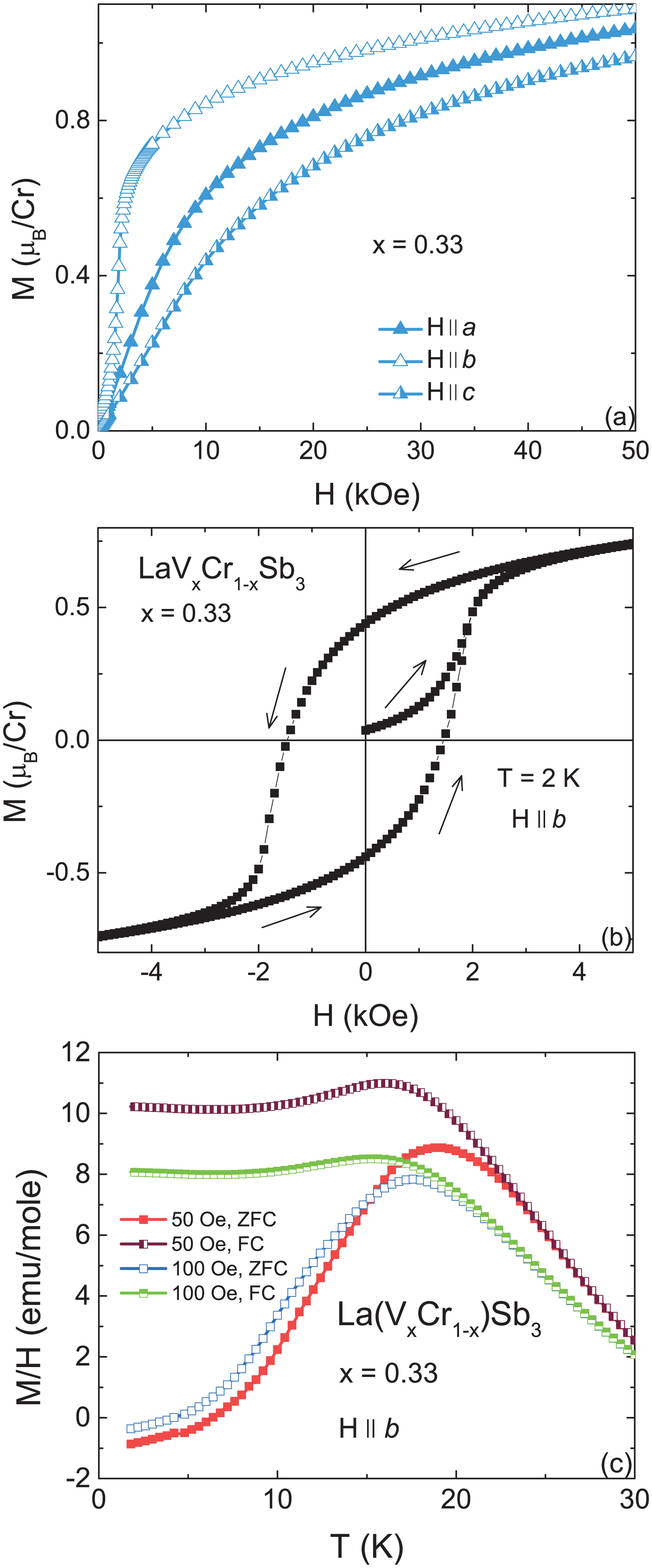}}
\caption{\label{fig:Sb-MH_33} (a) Zero-field-cooled (ZFC) anisotropic field-dependent magnetization isotherms taken at 2 K. (b) Hysteresis loop measured at 2 K with \textbf{H} $\parallel b$. The applied magnetic field changes from zero to 50 kOe, to -50 kOe, and then up to 50 kOe again. The arrows indicate the directions of the field sweeping. (c) Zero-field-cooled (ZFC) and field-cooled (FC) temperature dependence of the magnetic susceptibility taken at 50 and 100 Oe with \textbf{H} $\parallel b$ for the La(V$_x$Cr$_{1-x}$)Sb$_3$ ($x$ = 0.33).}
\end{center}
\end{figure}

\begin{figure}
\begin{center}
\resizebox*{8.5cm}{!}{\includegraphics{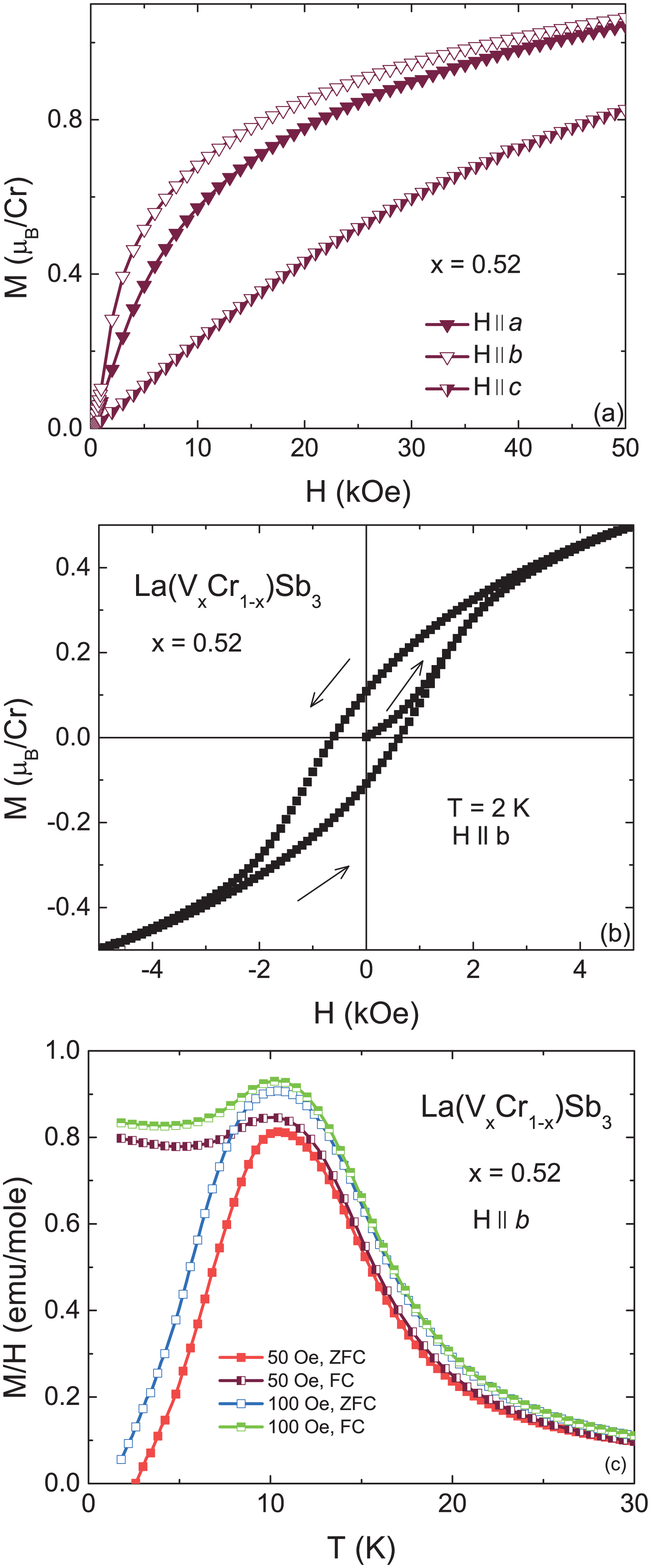}}
\caption{\label{fig:Sb-MH_52} (a) Zero-field-cooled (ZFC) anisotropic field-dependent magnetization isotherms taken at 2 K. (b) Hysteresis loop measured at 2 K with \textbf{H} $\parallel b$. The applied magnetic field changes from zero to 5 kOe, to -5kOe, and then up to 5 kOe again. The arrows indicate the directions of the field sweeping. (c) Zero-field-cooled (ZFC) and field-cooled (FC) temperature dependence of the magnetic susceptibility taken at 50 and 100 Oe with \textbf{H} $\parallel b$ for the La(V$_x$Cr$_{1-x}$)Sb$_3$ ($x$ = 0.52).}
\end{center}
\end{figure}

\begin{figure}
\begin{center}
\resizebox*{8.5cm}{!}{\includegraphics{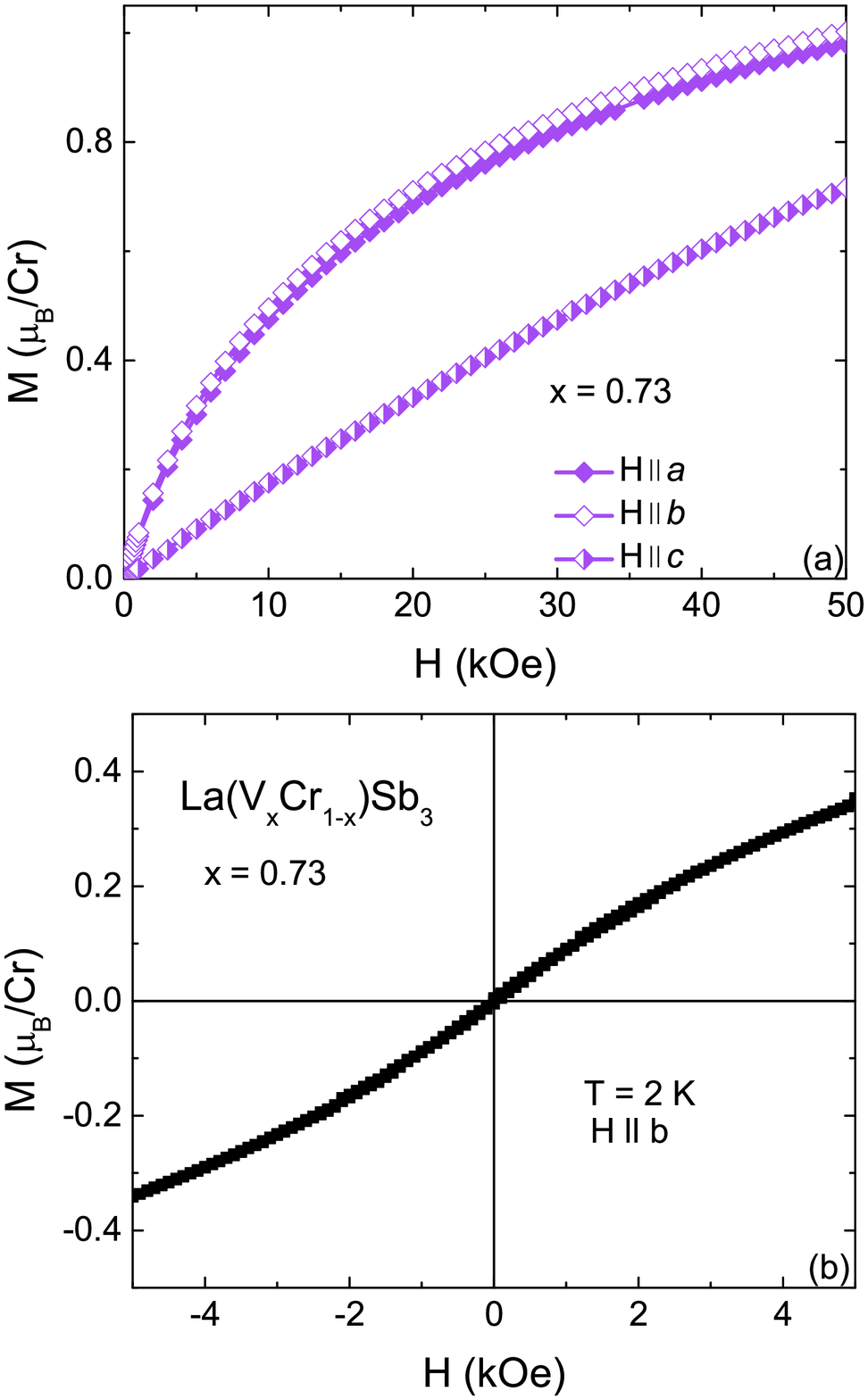}}
\caption{\label{fig:Sb-MH_73} (a) Zero-field-cooled (ZFC) anisotropic field-dependent magnetization isotherms taken at 2 K. (b) Hysteresis loop measured at 2 K with \textbf{H} $\parallel b$ for the La(V$_x$Cr$_{1-x}$)Sb$_3$ ($x$ = 0.73). The applied magnetic field changes from zero to 50 kOe, to -50 kOe, and then up to 50 kOe again.}
\end{center}
\end{figure}

To better understand the magnetic state of the V-doped compounds, systematic magnetization measurements were also performed. Figures \ref{fig:Sb-MH_0} -- \ref{fig:Sb-MH_73} present magnetization isotherms, hysteresis loops, zero-field-cooled (ZFC) and field-cooled (FC) $M(T)$ for selected compounds. Figure \ref{fig:Sb-MH_0} (a) shows the anisotropic, ZFC, magnetization isotherms for LaCrSb$_3$ measured at $T$ = 2 K. The magnetization shows clear ferromagnetic behavior -- spontaneous spin alignment in both $b$ and $c$ directions as the applied field increases from zero. At $T$ = 2 K, in the ordered state, the magnetization is anisotropic, with $M_b > M_c > M_a$. The value of the magnetization measured at 50 kOe in the $b$ direction is taken as the saturated moment ($\mu_{\rm S}$). For $x$ = 0,  $\mu_{\rm S}$ is about 1.61 $\mu_{\rm B}$ per Cr, consistent with the reported value \cite{Jackson-PRB01, Granado-PRL02}. Figure \ref{fig:Sb-MH_0} (b) shows the hysteresis loop of LaCrSb$_3$ measured at 2 K for \textbf{H} $\parallel b$. The spontaneous spin alignment can be clearly seen, whereas hysteresis can hardly be observed. This probably suggests that LaCrSb$_3$ is a soft ferromagnet. Compared to the V-doped compounds (see below), LaCrSb$_3$ exhibits negligible pinning effect that is associated with the disorder induced by substitution.

Similar magnetization isotherms can be observed for $x$ = 0.22, see in fig. \ref{fig:Sb-MH_22} (a). The $b$-axis can still be identified as the easy axis and $M_a$ has the lowest value of all three directions, however, the differences in magnetization between different directions becomes slightly less obvious. The saturated moment is 1.37 $\mu_{\rm B}$/Cr for \textbf{H} $\parallel b$. The hysteresis loop for $x$ = 0.22 is shown in fig. \ref{fig:Sb-MH_22} (b). Clear hysteresis can be observed, and the coercivity is about 1.63 kOe. At low fields, the magnetization in the virgin curve, instead of showing spontaneous spin alignment, rises slowly with the applied field. This is probably caused by domain pinning effects. In addition, a discrepancy can be seen in the low-field magnetization for \textbf{H} $\parallel b$ between fig. \ref{fig:Sb-MH_22} (a) and fig. \ref{fig:Sb-MH_22} (b). It is possibly due to the remnant field in a superconducting magnet giving rise to different virgin curve starting points. Based on the behavior of the field-dependent magnetization, it is evident that La(V$_x$Cr$_{1-x}$)Sb$_3$ ($x$ = 0.22) possesses a ferromagnetic state at low temperatures. Figure \ref{fig:Sb-MH_22} (c) shows the ZFC and FC magnetization as a function of temperature. The measurements were performed at 50 and 100 Oe with \textbf{H} $\parallel b$. As can be seen, the FC $M/H$ increases dramatically upon cooling and continuously rising at low temperatures, indicating the existence of a ferromagnetic state. The complex feature in the ZFC $M(T)/H$ at low temperatures is probably due to domain pinning effects. Hence, with increasing V substitution up till $x$ = 0.22, the La(V$_x$Cr$_{1-x}$)Sb$_3$ series maintains a ferromagnetic state at low temperatures.

Starting from $x$ = 0.33, two major differences can be found in the magnetization isotherms, as shown in fig. \ref{fig:Sb-MH_33} (a) and fig. \ref{fig:Sb-MH_52} (a). First of all, a spontaneous spin alignment can not be observed for any direction of applied field measured. The magnetization for \textbf{H} $\parallel b$ rises much slower as field increases (compared with the case of $x$ = 0), whereas $M_a$ and $M_c$ seem to show rather broad shoulders. With the increased V concentration, even $M_b$ shows a shoulder-like feature, and no saturation can be observed. These data might suggest that the magnetic state for $x \geq$ 0.33 in the La(V$_x$Cr$_{1-x}$)Sb$_3$ series is no longer ferromagnetic. The second difference found is the change of anisotropy. Although the $b$-axis is still the easy axis, $M_a$ is larger than $M_c$ for $x \geq$ 0.33. Hysteresis can still be observed for $x$ = 0.33 and 0.52, as shown in fig. \ref{fig:Sb-MH_33} (b) and fig. \ref{fig:Sb-MH_52} (b). However the coercivity decreases as the V concentration increases. It drops to 1.47 kOe for $x$ = 0.33 and 0.58 kOe for $x$ = 0.52.

Figures \ref{fig:Sb-MH_33} (c) and \ref{fig:Sb-MH_52} (c) also present the ZFC and FC magnetization as a function of temperature measured at 50 and 100 Oe with \textbf{H} $\parallel b$. Besides the initial increase in both ZFC and FC $M(T)/H$ upon cooling, a local maximum can be observed in both of the ZFC and FC curves. This feature is more obvious for the higher V-doped compound, $x$ = 0.52 (fig. \ref{fig:Sb-MH_52} (c)). As is shown, the ZFC and FC $M(T)/H$ are split at low temperatures, and the ZFC curves exhibit a dramatic decrease as $T$ decreases. These might imply some degree of frustration which leads to some form of cluster or spin-glass state \cite{Mydosh-93}. It is possible that with the continuous suppression of ferromagnetism, the magnetic state in this series evolves into a new magnetic state, which is often observed in the local magnetic moment systems \cite{Mydosh-93, Buschow90}. 

As vanadium content increases, the moment along $a$-axis continuously gets closer to $M_b$. As shown in fig. \ref{fig:Sb-MH_73} (a), $M_a$ and $M_b$ becomes almost identical for $x$ = 0.73. At 2 K, the magnetization gradually increases with the increasing field, exhibiting no feature of spontaneous spin alignment. No saturation or hysteresis can be observed for $x$ = 0.73 (fig. \ref{fig:Sb-MH_73} (b)). It is evident that La(V$_x$Cr$_{1-x}$)Sb$_3$ series does not possess a ferromagnetic order above 2.0 K for $x >$ 0.73.

\begin{figure}
\begin{center}
\resizebox*{8.5cm}{!}{\includegraphics{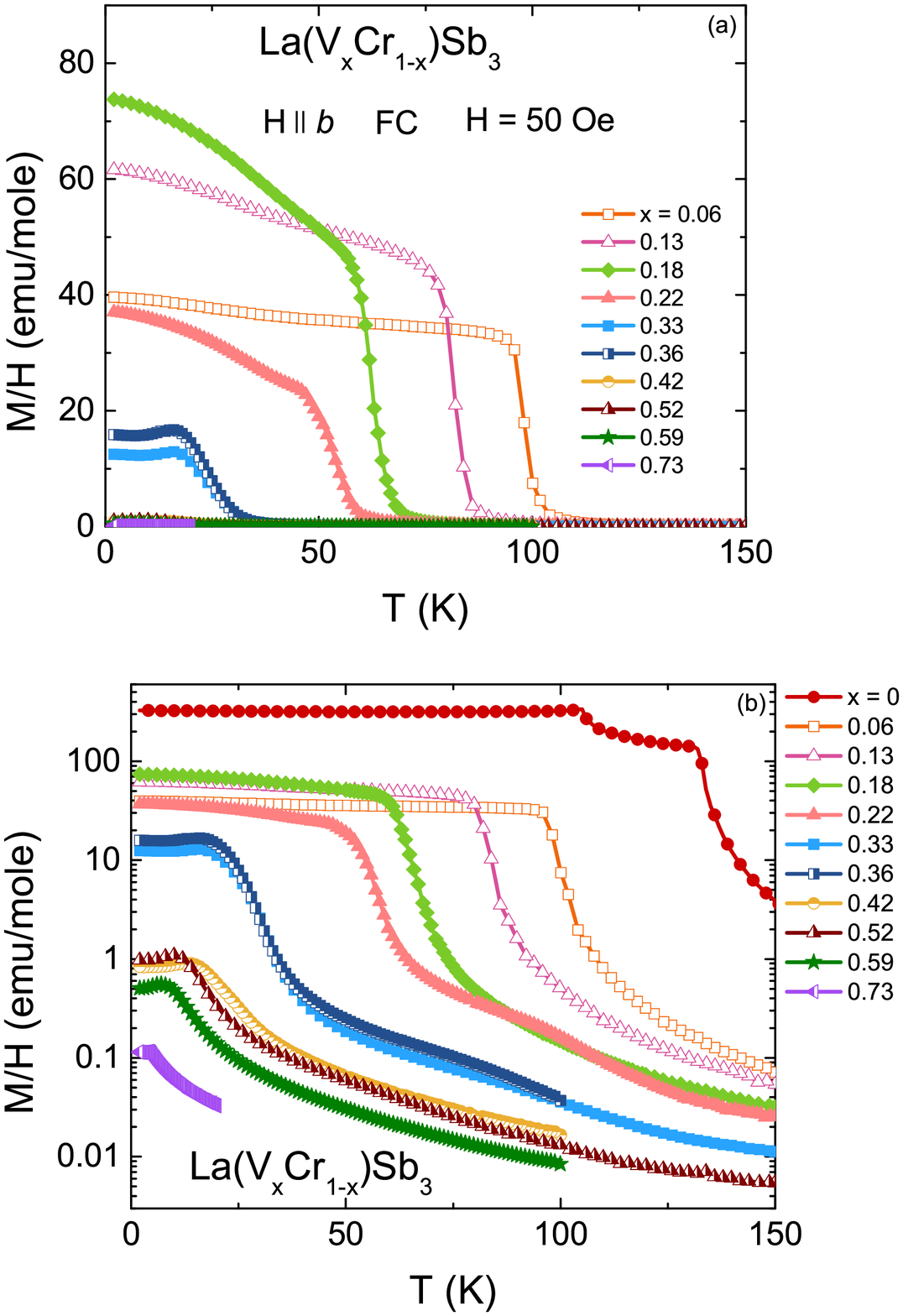}}
\caption{\label{fig:Sb-MT} (a) FC magnetization as a function of temperature for La(V$_x$Cr$_{1-x}$)Sb$_3$ ($x$ = 0.06 -- 0.73) at 50 Oe with \textbf{H} $\parallel b$. (b) FC magnetization as a function of temperature for $x$ = 0 -- 0.73 presented in a semi-log plot.}
\end{center}
\end{figure}

Given that the magnetization along $b$-axis shows typical ferromagnetic behavior for the lower V-doped compounds, and given that the $b$-axis is the easy axis in almost the whole $x$ range ($0 \leq x \leq 0.73$), in the following, our study of the evolution of the magnetic state in the La(V$_x$Cr$_{1-x}$)Sb$_3$ system is presented with the magnetization data along $b$-axis. The temperature-dependent FC magnetization curves of the La(V$_x$Cr$_{1-x}$)Sb$_3$ ($x$ = 0.06 -- 0.73) series, with \textbf{H} $\parallel b$ at 50 Oe are shown in fig. \ref{fig:Sb-MT} (a) and (b). Magnetization for $x$ = 0.06 -- 0.36 shows the expected rapid increase of the magnetization as well as the saturation at low temperatures (fig. \ref{fig:Sb-MT} (a)). The Curie temperature decreases as V concentration increases, and the transition shifts to lower temperature, as can be clearly seen in fig. \ref{fig:Sb-MT} (b). With increasing amounts of V substituted for Cr, from $x$ = 0.42, the temperature-dependent magnetic susceptibility starts deviating from the ferromagnetic behavior (fig.\ref{fig:Sb-MT}). As the temperature decreases, the magnetization rises in a much slower manner compared with the lower V-doped compounds, and a local maximum at low temperatures can be observed. It can be inferred that instead of the ferromagnetic state, a new magnetic state may emerge for higher V-doped compounds.

\begin{figure}
\begin{center}
\resizebox*{8.5cm}{!}{\includegraphics{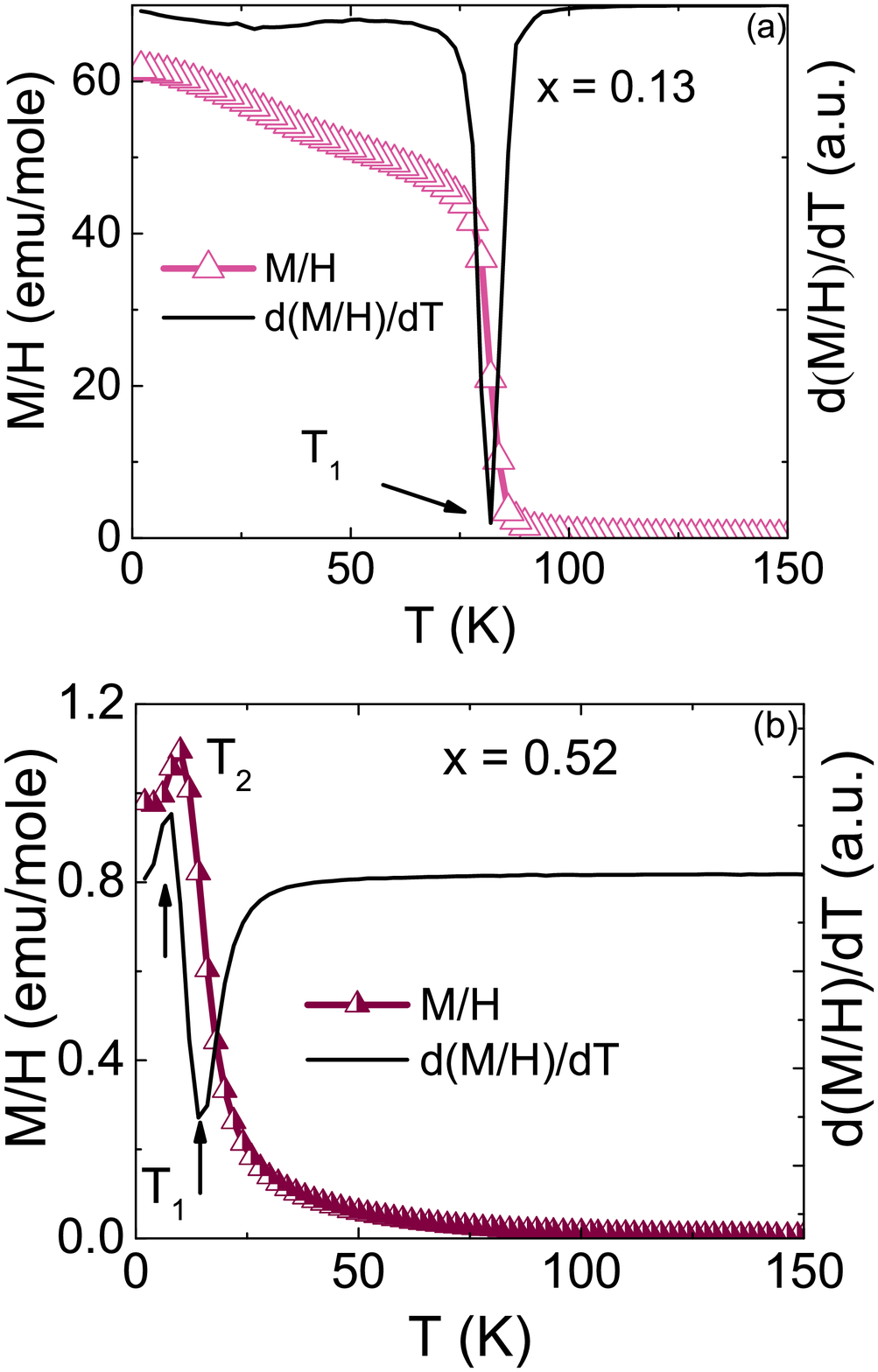}}
\caption{\label{fig:Sb-Criterion} (a) The temperature dependence of the magnetization and $d(M/H)/dT$ for $x$ = 0.13, and the arrow indicates the criterion used to determine the transition temperature $T_1$. (b) The temperature dependence of the magnetization and $d(M/H)/dT$ for $x$ = 0.52, and the arrow indicates the criteria used to determine the transition temperature $T_1$ and $T_2$.}
\end{center}
\end{figure}

The Curie temperature can be estimated by $d(M/H)/dT$ for low values of applied field ($H$ = 50 Oe in this case), as indicated by the arrow in fig. \ref{fig:Sb-Criterion} (a), where the transition temperature $T_1$ is determined by the sharp anomaly in $d(M/H)/dT$. The obtained $T_1$-values are listed in Table \ref{tab:Sb-Temperature}. The transition temperature systematically decrease to lower temperatures as the V-concentration increases, from 133 K for $x$ = 0 to 24 K for $x$ = 0.36 with \textbf{H} $\parallel b$. On the other hand, for higher V-doped compounds ($x >$ 0.33), two features can be observed in $d(M/H)/dT$. Besides the minimum point, the maximum point in $d(M/H)/dT$ is chosen as the criterion to characterize the transition temperature of a potential new magnetic state (shown in fig.\ref{fig:Sb-MT} (b)). Again the obtained transition temperatures $T_1$ and $T_2$ are listed in Table \ref{tab:Sb-Temperature}.

\begin{figure}
\begin{center}
\resizebox*{8.5cm}{!}{\includegraphics{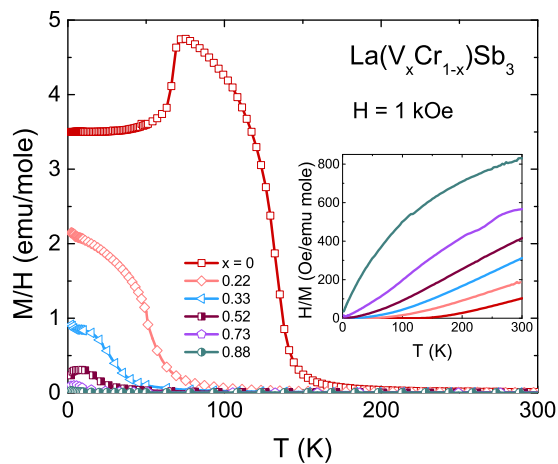}}
\caption{\label{fig:Sb-Eff} Polycrystalline averaged $M/H$ vs. $T$ for La(V$_x$Cr$_{1-x}$)Sb$_3$ ($x$ = 0, 0.22, 0.33, 0.52, 0.73 and 0.88) measured at $H$ = 1 kOe. Inset: inverse magnetic susceptibility as a function of temperature.}
\end{center}
\end{figure}

The polycrystalline average of $M/H$ measured at 1 kOe is shown in fig. \ref{fig:Sb-Eff}. It is obtained by $\chi_{ave}$ = $\frac{1}{3}$ $(\chi_{a}+\chi_{b}+\chi_{c})$. A modified Curie-Weiss law with inclusion of a temperature-independent term $\chi_{\rm 0}$: $\chi_{\rm ave}$ = $\chi_{\rm 0}$ + $C/(T-\theta_{\rm poly})$, was used to fit the magnetic susceptibility, where $\theta_{\rm poly}$ is the Curie-Weiss temperature estimated by the polycrystalline averaged data and $C$ is the Curie constant. Considering the accuracy of measuring sample's mass, the values of the effective moments in this series are accurate to $\pm$10$\%$. The fitting parameters $\chi_{\rm 0}$ and the calculated $\mu_{\rm eff}$ and $\theta_{\rm poly}$ are summarized in Table \ref{tab:Sb-Temperature}. For $x$ = 0, $\mu_{\rm eff}$ is found to be about 3.9 $\mu_{\rm B}$/Cr, close to the calculated value for  Cr$^{3+}$: 3.8 $\mu_{\rm B}$, and is consistent with the reported value \cite{Jackson-PRB01}. As shown in Table \ref{tab:Sb-Temperature}, the effective moment gradually decreases as the V concentration increases. However, $\mu_{\rm eff}$ does not approach zero for some critical $x$ value, or even as $x$ gets close to 1, unlike the ferromagnetic system dominated by solely itinerant moments \cite{Rhodes-PRSLA63}. Instead, $\mu_{\rm eff}$ falls to 2.9 $\mu_{\rm B}$/Cr for $x$ = 0.88, which is close to Cr$^{4+}$: 2.8 $\mu_{\rm B}$. It is possible that the Cr ion in the La(V$_x$Cr$_{1-x}$)Sb$_3$ compounds has a valence changed in compounds with higher V substitution. In addition, it is found that the Curie-Weiss temperature, $\theta_{\rm poly}$ in this series are all positive, indicating the ferromagnetic interaction as the dominant interaction in these compounds. Also the fact that $\theta_{\rm poly}$ decreases as $x$ increases implies the ferromagnetic interaction is weakened by the V substitution.


\begin{figure}
\begin{center}
\resizebox*{8.5cm}{!}{\includegraphics{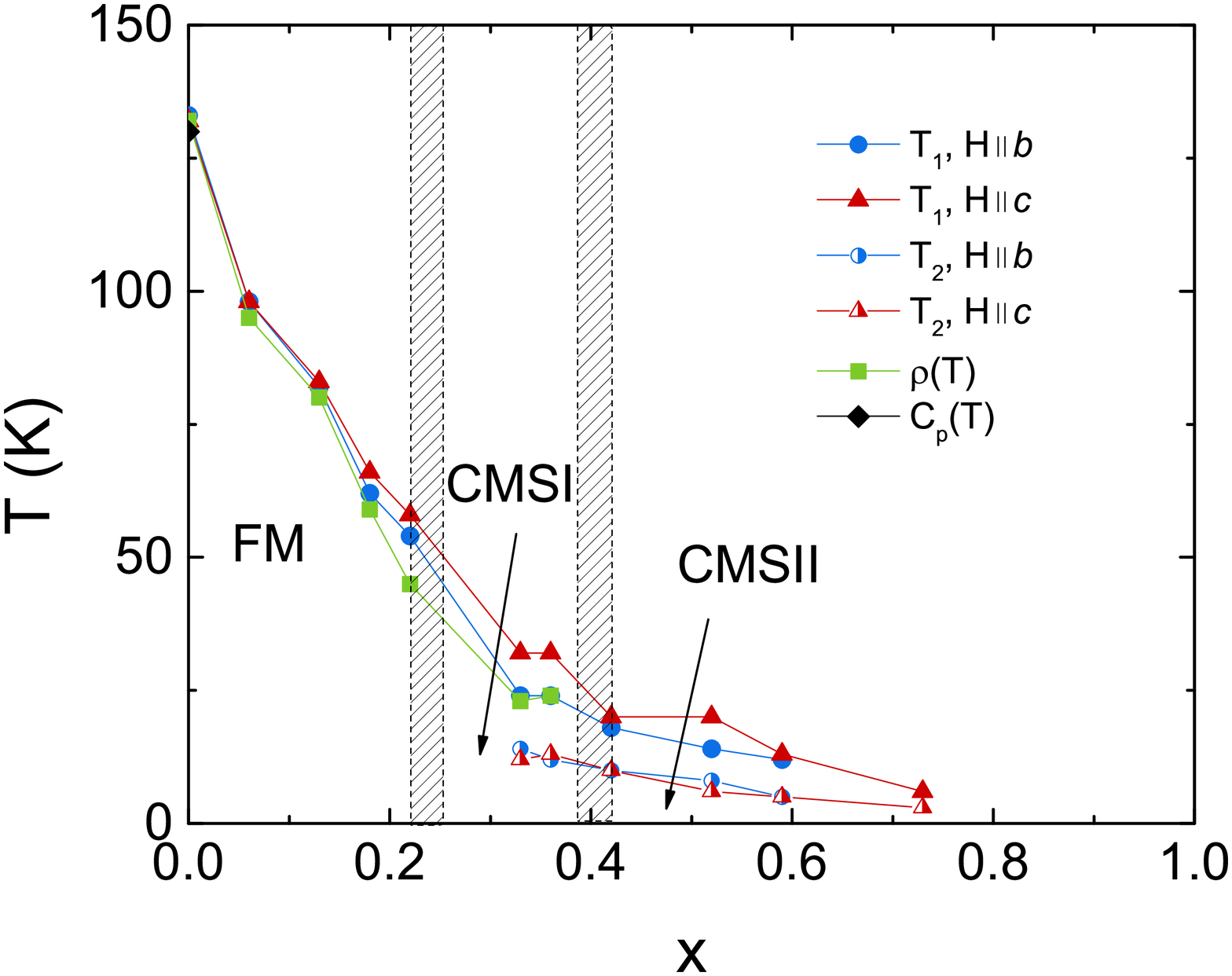}}
\caption{\label{fig:Sb-PDall} $x$-dependent transition temperatures for La(V$_x$Cr$_{1-x}$)Sb$_3$ determined by anisotropic $M(T)$, $\rho(T)$ and $C_{\rm p}(T)$ measurements. The dashed lines outline three potential regions of the low temperature magnetic behavior: ferromagnetic state (FM), complex magnetic state I (CMSI) and complex magnetic state II (CMSII).}
\end{center}
\end{figure}

We have been able to suppress the ferromagnetism in the La(V$_x$Cr$_{1-x}$)Sb$_3$ series via chemical substitution. The ordering temperatures inferred from low field magnetization, resistivity and specific heat measurements are summarized in Table \ref{tab:Sb-Temperature}. A phase diagram of $x$-dependent transition temperature for La(V$_x$Cr$_{1-x}$)Sb$_3$ is assembled in fig. \ref{fig:Sb-PDall}. For $x \leqslant$ 0.36, the transition temperatures that are determined by different measurements and different orientations are fairly consistent. For $x \geqslant$ 0.42, only the points inferred from the magnetization data are shown in the phase diagram, since other measurements do not manifest clear anomalies. As can be seen, for the La(V$_x$Cr$_{1-x}$)Sb$_3$ series, ferromagnetism can be clearly observed for $x$ up to 0.22, and the ferromagnetic transition temperature is suppressed monotonically by the V substitution: $T_{\rm C}$ = 133 K for $x$ = 0, and $T_{\rm C}$ = 52 K for $x$ = 0.22 (based on the low field $M(T)$ data with \textbf{H} $\parallel b$). If $T_1$ is used as a criterion for determining the transition to a low-temperature magnetic state for the whole series, it seems that the magnetic transition temperature gets gradually suppressed by V substitution and drops below our base-temperature of 2.0 K. If $T_2$ is used as the criterion for higher V-doped samples, then considering the features observed in $M (H)$, $M(T)/H$ and $\rho (T)$ for $x$ = 0.33 and 0.36, it seems that $x$ = 0.33 -- 0.36 is a region for the system to transition from the ferromagnetic state to a new magnetic state. Similar phenomena have also been observed in the LiHo$_x$Y$_{1-x}$F$_4$ family: for 0.25 $\leq x \leq$ 0.5, the system is claimed to be in a ``ferroglass" regime, where spin glass and ferromagnetic phase coexist \cite{Ancona-Torres-PRL08, Gingras-JPCS11}. In fig. \ref{fig:Sb-PDall} we identify three potential regions of the low temperature magnetic behavior: (1) ferromagnetic state (FM) for 0 $\leq x \leq$ 0.22, (2) complex magnetic state I (CMSI) for  0.22 $\leq x \leq$ 0.42, and (3) complex magnetic state II (CMSII) for $x \geq$ 0.42. Further investigations are needed to determine the exact concentration at which the transition occurs. In the region of CMSII, given that i) $M (T)$ exhibits a local maximum at low temperature, and ii) field-cooled and zero-field-cooled $ M (T)$ deviate from each other, it is possible that this new magnetic ground state is a complex glassy state. As the V concentration reaches even higher, $ x \geq 0.88$, no magnetic ordering can be observed and the system shows paramagnetic behavior down to our base temperature of 2.0 K.

Based on our data, the La(V$_x$Cr$_{1-x}$)Sb$_3$ system has a phase diagram consistent with dominantly local moment like behavior of Cr. The progression from well defined magnetic ordering to complex magnetic state, to something that may be glassy state and gradually has a freezing temperature drop toward zero is similar to what has been found for local moment systems such as (Tb$_x$Y$_{1-x}$)Ni$_2$Ge$_2$ \cite{Wiener-PRB00}. At no point were features consistent with a quantum critical point observed.

\begin{table}
  \tbl{Summarized effective moment $\mu_{\rm eff}$, saturated moment $\mu_{\rm B}$, polycrystalline averaged Curie-Weiss temperature $\theta_{\rm poly}$, transition temperatures $T_1$, $T_2$ estimated from low field M(T) measurements with \textbf{H} $\parallel b$ and \textbf{H} $\parallel c$, and Curie temperature $T_{\rm C}$ estimated from $\rho(T)$ and $C_{\rm p}(T)$ measurements for La(V$_x$Cr$_{1-x}$)Sb$_3$ ($x$ = 0 -- 0.88).}
{\begin{tabular}{|l|l|l|l|l|l|l|l|l|l|l|}\toprule
$x$	&	$T_{\rm C}$ &	$T_1$  &	$T_2$ &	$T_1$ 	&	$T_2$ 	&	$T_{\rm C}$ &	$\theta_{\rm poly}$	&	$\chi_0$  	&	$\mu_{\rm eff}$&$\mu_{\rm S}$ \\
 
&	(K)	&	(K)	&	(K)	&	(K)	&	(K)	&	(K)	&	(K) 	&	($10^{-4}$ emu/mole)	&($\mu_B$/Cr) &($\mu_B$/Cr)\\
 
&	$d\rho/dT$	&	\textbf{H} $\parallel b$	&	\textbf{H} $\parallel b$	&	\textbf{H} $\parallel c$	&	\textbf{H} $\parallel c$	&	$C_p(T)$	&	 	&		&&\\
\colrule
 0	    &	132	&	133	&		&	132	&		&	132	&	141	&	-40	&	3.9&1.61	\\
0.06	&	95	&	98	&		&	98	&		&		&		&		&	&	\\
0.13	&	80	&	82	&		&	83	&		&		&		&		&	&	\\
0.18	&	59	&	62	&		&	66	&		&		&		&		&	&	\\
0.22	&	45	&	54	&		&	58	&		&		&	90	&	0	&	3.7&1.37	\\
0.33	&	23	&	24	&	14 	&	32	&	12 	&		&	70	&	-5	&	3.2&	\\
0.36	&	24	&	24	&	12 	&	32	&	13 	&		&		&		&		&\\
0.42	&		&	18	&	10	&	20	&	10	&		&		&		&		&\\
0.52	&	 	&	14	&	8	&	20	&	6	&		&	52	&	1	&	3.1	&\\
0.59	&	 	&	12	&	5	&	13	&	5	&		&		&		&		&\\
0.73	&	 	&	 	&		&	6	&	3	&		&	34	&	6	&	2.9	&\\
0.88	&	 	&	 	&		&	 	&		&		&	0.1	&	8	&	2.9	&\\
   \botrule
  \end{tabular}}
\label{tab:Sb-Temperature}
\end{table}

\section{\label{sec:level1d}Discussion and Conclusions}

Our efforts to suppress ferromagnetism in LaCrSb$_3$ started with applications of pressures up to 5.3 GPa. As seen in fig. \ref{fig:Sb-Pressure}, the ferromagnetic ordering temperature, $T_{\rm C}$, is essentially insensitive to $p <$ 5.3 GPa. On the contrary, the feature of spin reorientation evolves systematically and vanishes as the applied pressure increases.

Given that we could not suppress the ferromagnetism in the LaCrSb$_3$ compound by applying pressure, we evaluated the potential for quantum critical behavior by using chemical substitution as an alternative tuning parameter. The growth of single crystalline La(V$_x$Cr$_{1-x}$)Sb$_3$ samples has allowed for the detailed study of the anisotropic properties, the determination of the easy axis as well as the estimate of the effective moment. In addition, careful chemical analysis was performed to determine the precise concentration of this doped system. This offers a certain understanding of chemical substitution effect on the suppression of the ferromagnetism and the evolution of the magnetic state in this system.

\begin{figure}
\begin{center}
\resizebox*{8.5cm}{!}{\includegraphics{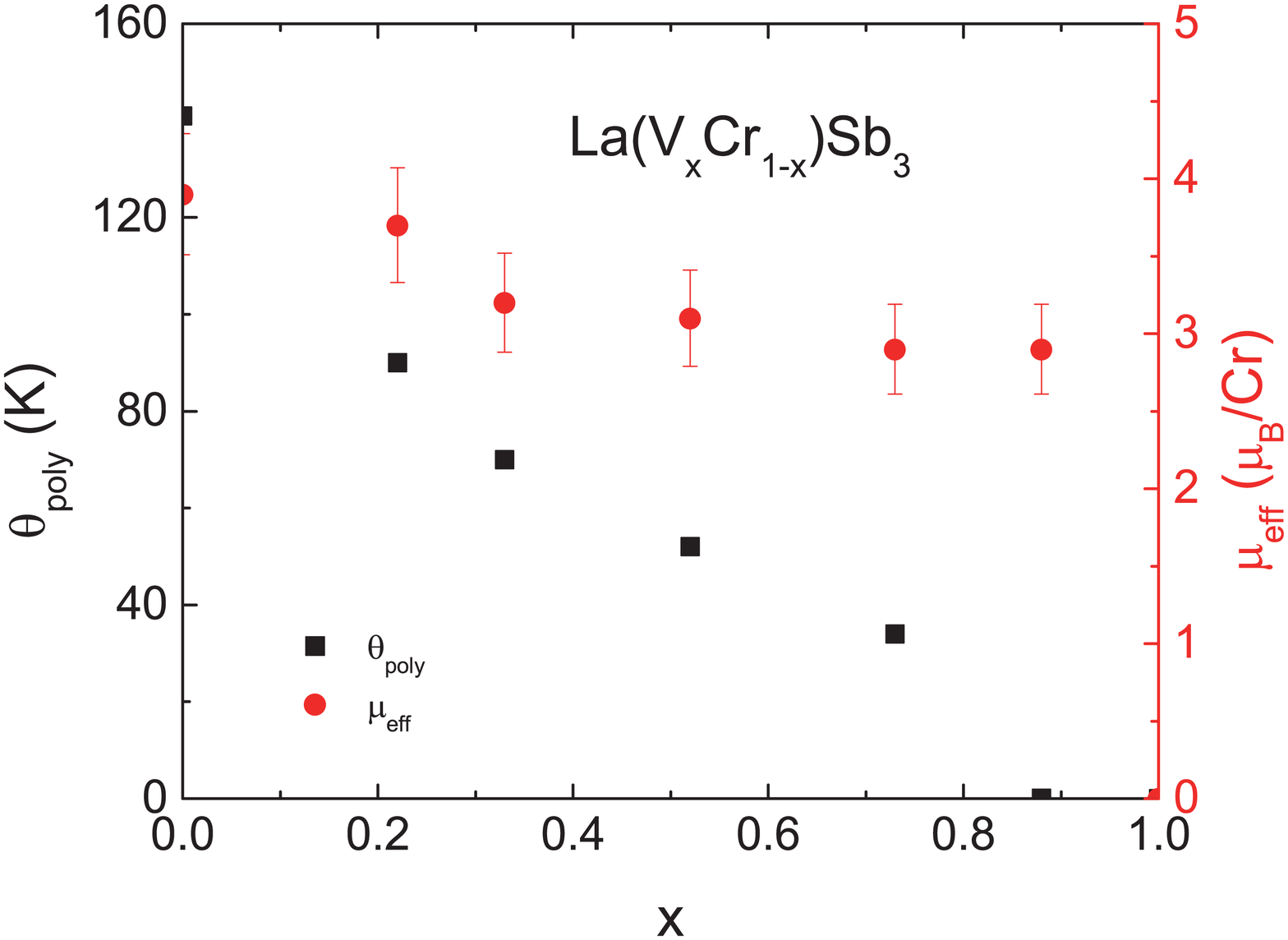}}
\caption{\label{fig:Sb-Moment-T} The Curie-Weiss temperature $\theta_{\rm poly}$ and the effective moment $\mu_{\rm eff}$ per Cr as a function of $x$ for La(V$_x$Cr$_{1-x}$)Sb$_3$.}
\end{center}
\end{figure}

\begin{figure}
\begin{center}
\resizebox*{8.5cm}{!}{\includegraphics{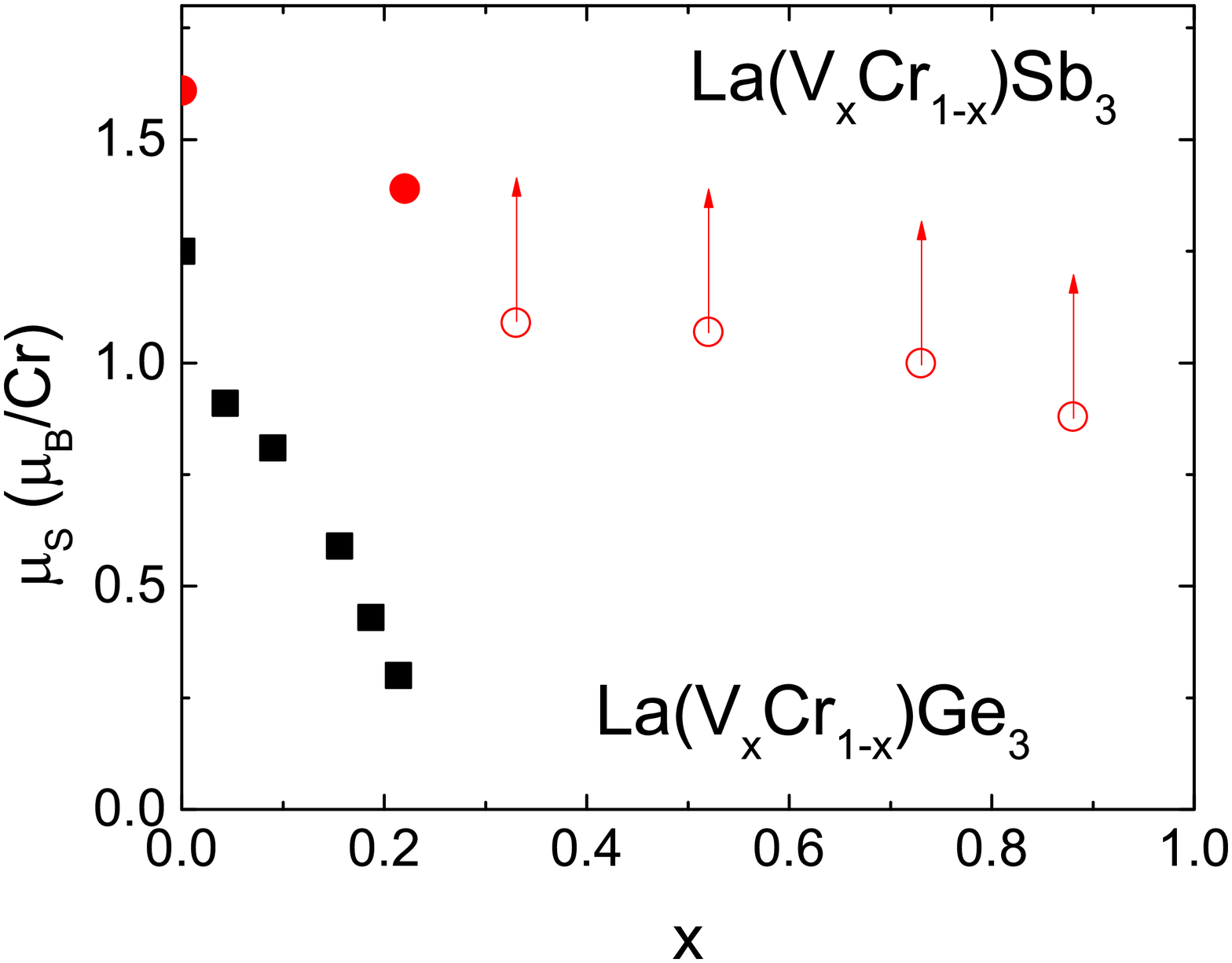}}
\caption{\label{fig:Sb-Ms} The saturated moment $\mu_{S}$ as a function of $x$ for the La(V$_x$Cr$_{1-x}$)Ge$_3$ and La(V$_x$Cr$_{1-x}$)Sb$_3$ series. Note: for La(V$_x$Cr$_{1-x}$)Sb$_3$ compounds with $x >$ 0.22, the saturated moment is replaced by the value of magnetization at 50 kOe with field along the $b$-axis. Data on La(V$_x$Cr$_{1-x}$)Ge$_3$ are from Ref. \cite{Lin-Ge}.}
\end{center}
\end{figure}

\begin{figure}
\begin{center}
\resizebox*{8.5cm}{!}{\includegraphics{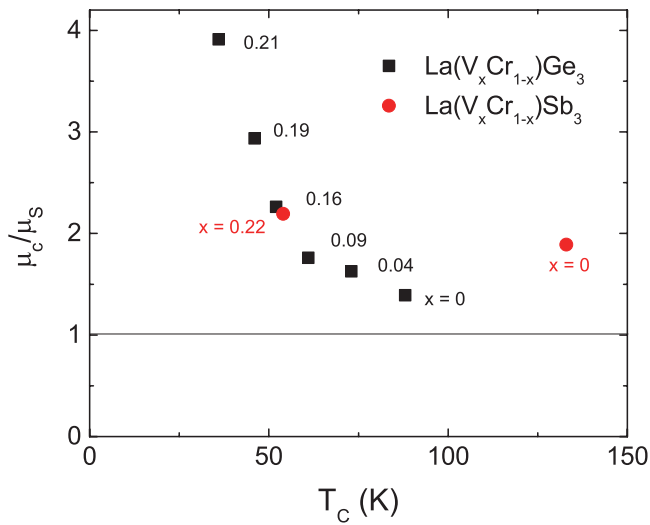}}
\caption{\label{fig:Sb-RWR} The Rhodes-Wohlfarth ratio $\mu_{\rm c}$/$\mu_{\rm S}$ as a function of Curie temperature $T_{\rm C}$ for the La(V$_x$Cr$_{1-x}$)Ge$_3$ and La(V$_x$Cr$_{1-x}$)Sb$_3$ series. Data on La(V$_x$Cr$_{1-x}$)Ge$_3$ are from Ref. \cite{Lin-Ge}.}
\end{center}
\end{figure}

The estimated Curie-Weiss temperature $\theta_{\rm poly}$ and the effective moment $\mu_{\rm eff}$ per Cr as a function of the V-concentration $x$ are plotted in fig. \ref{fig:Sb-Moment-T}. It is clearly seen that $\theta_{\rm poly}$ decreases monotonically from about 141 K to almost zero as $x$ increases, implying that the ferromagnetic interaction is suppressed by the V substitution. The effective moment also decreases as $x$ increases, but in a more subtle way. Instead of quickly approaching to zero as seen in the suppression of ferromagnetism in the itinerant ferromagnets \cite{Stewart-RMP84, Stewart-RMP01, Stewart-RMP06, Taufour-PRL10}, $\mu_{\rm eff}$ per Cr varies slightly as $x$ increases, from 3.9 $\mu_{\rm B}$ for $x$ = 0 to 2.9 $\mu_{\rm eff}$ for $x$ = 0.88. Whereas the former value is close to the theoretical effective moment of Cr$^{3+}$, the latter one is close to Cr$^{4+}$’. Hence, the decrease of $\mu_{\rm eff}$ is more likely as a consequence of the valence change due to the V substitution than manifesting an itinerant-moment behavior. 

The saturated moment values for the La(V$_x$Cr$_{1-x}$)Sb$_3$ compounds are presented in fig. \ref{fig:Sb-Ms}. $\mu_{\rm S}$ shows a slight decrease with the increase of the V concentration, from 1.61 $\mu_B$/Cr for $x$ = 0 to 1.37 $\mu_B$/Cr for $x$ = 0.22. For higher V-doped compounds ($x >$ 0.22), the La(V$_x$Cr$_{1-x}$)Sb$_3$ systems does not have a ferromagnetic state at low temperatures, hence, $\mu_{\rm S}$ is replaced by the value of magnetization at 50 kOe with field along the $b$-axis. The arrows in fig. \ref{fig:Sb-Ms} imply that $M$($H$ = 50 kOe) is the lower limit of the possible saturated moment of the higher V-doped compounds. It is clear that Cr's moment stays well above zero as the V concentration approaches to 1.

To learn more about Cr's magnetic moment, we can compare the saturated moments of both La(V$_x$Cr$_{1-x}$)Ge$_3$ and La(V$_x$Cr$_{1-x}$)Sb$_3$ series (as shown in fig. \ref{fig:Sb-Ms}). It can be clearly seen that $\mu_{\rm S}$ of the La(V$_x$Cr$_{1-x}$)Ge$_3$ compounds quickly decreases as the V concentration increase, that is associated with the itinerant magnetism \cite{Lin-Ge}. For the La(V$_x$Cr$_{1-x}$)Sb$_3$ series, $\mu_{\rm S}$ has a slower decreasing rate as the V concentration increases. This possibly implies that the magnetic moment associated with Cr in the La(V$_x$Cr$_{1-x}$)Sb$_3$ compounds is mainly of local character. Based on the values of $\mu_{\rm eff}$ and $\mu_{\rm S}$ obtained, one can calculate the Rhodes-Wolfarth ratio (RWR) \cite{Rhodes-PRSLA63}, seen in fig. \ref{fig:Sb-RWR}. According to Rhodes and Wolfarth, RWR = $\mu_{\rm c}$/$\mu_{\rm S}$, where $\mu_{\rm c}$ is related to the number of moment carriers, and can be obtained from $\mu_{\rm c}$($\mu_{\rm c}$+1)=$\mu_{\rm eff}^2$. While RWR = 1 is an indication of localized magnetism, larger RWR values suggest the existence of itinerant ferromagnetism. In our case, RWR equals to $\simeq$ 1.9 for $x$ = 0 and $\simeq$ 2.2 for $x$ = 0.22. Although this seems suggesting the ferromagnetism is itinerant, the change of RWR as a function of $T_{\rm C}$ shows very different behavior compared with the La(V$_x$Cr$_{1-x}$)Ge$_3$ system \cite{Lin-Ge} (seen in fig. \ref{fig:Sb-RWR}) and the original Rhodes-Wohlfarth plot \cite{Rhodes-PRSLA63}. Therefore, unlike La(V$_x$Cr$_{1-x}$)Ge$_3$, La(V$_x$Cr$_{1-x}$)Sb$_3$ is less likely to be dominated solely by itinerant magnetic moments. Probably the magnetism in this family is of predominantly local character. 

\section*{Acknowledgements}

We thank W. E. Straszheim for his assistance with the elemental analysis of the samples. We appreciate help of E. Colombier and A. Kaminski in setting up the ruby fluorescence system. We also would like to thank H. Hodovanets for assisting in Laue diffraction measurement. This work was carried out at the Iowa State University and supported by the AFOSR-MURI grant No. FA9550-09-1-0603 (X. Lin, V. Taufour and P. C. Canfield). S. L. Bud'ko was supported by the U.S. Department of Energy, Office of Basic Energy Science, Division of Materials Sciences and Engineering. Part of this work was performed at Ames Laboratory, US DOE, under Contract No. DE-AC02-07CH11358.

\end{document}